\documentclass[prd,aps,groupedaddress,
showpacs,preprintnumbers,
twocolumn,%
nofootinbib
]{revtex4}%
\usepackage{graphicx}
\usepackage{amsmath}
\usepackage{bm}
\usepackage{epsfig}
\usepackage{amssymb}

\begin{document}
\title{Factorization of low-energy gluons in exclusive processes}
\author{Geoffrey~T.~Bodwin}
\affiliation{High Energy Physics Division, Argonne National Laboratory,\\
9700 South Cass Avenue, Argonne, Illinois 60439, USA}
\author{Xavier \surname{Garcia i Tormo}}
\affiliation{High Energy Physics Division, Argonne National Laboratory,\\
9700 South Cass Avenue, Argonne, Illinois 60439, USA}
\affiliation{Department of Physics, University of Alberta,
Edmonton, Alberta, Canada T6G 2G7 \footnote{Current address}}
\author{Jungil~Lee}
\affiliation{Department of Physics, Korea University, Seoul 136-701, Korea}

\date{\today}

\preprint{\begin{tabular}{l}ANL-HEP-PR-09-6  \\Alberta Thy 08-10\end{tabular}}

\begin{abstract}
We outline a proof of factorization in exclusive processes, taking into
account the presence of soft and collinear modes of arbitrarily low
energy, which arise when the external lines of the process are taken on
shell. Specifically, we examine the process of $e^+e^-$ annihilation
through a virtual photon into two light mesons. In an intermediate step,
we establish a factorized form that contains a soft function that is
free of collinear divergences. In contrast, in soft-collinear
effective theory, the low-energy collinear modes factor most
straightforwardly into the soft function. We point out that the
cancellation of the soft function, which relies on the color-singlet
nature of the external hadrons, fails when the soft function contains
low-energy collinear modes.
\end{abstract}

\maketitle

\section{Introduction}

Factorization theorems are fundamental to modern calculations in 
QCD of the amplitudes for hard-scattering exclusive
hadronic processes. They allow one to separate contributions to the
amplitudes that involve states of high virtuality from those that
involve states of low virtuality. The former, short-distance
contributions can, by virtue of asymptotic freedom, be calculated in
perturbation theory, while the latter, long-distance contributions are
parametrized in terms of inherently nonperturbative matrix elements of
QCD operators in hadronic states.

States of low virtuality can arise from the emission of a soft gluon,
whose four-momentum components are all small, or from the emission of a
collinear gluon, whose four-momentum is nearly parallel to the
four-momentum of a gluon or light quark. In some discussions of
factorization that employ soft-collinear effective theory (SCET)
\cite{Bauer:2000yr,Bauer:2001yt,Bauer:2002nz} or diagrammatic methods
\cite{Bodwin:2008nf}, it is assumed that gluons can have no transverse
momentum components that are smaller than the QCD scale $\Lambda_{\rm
QCD}$. That is, gluons can have hard momentum, in which all components
are of order the hard-scattering scale $Q$, soft momentum, in which all
components are of order $\Lambda_{\rm QCD}$, or collinear momentum, in
which the transverse components are of order $\Lambda_{\rm QCD}$ and the
energy and longitudinal spatial component are much larger than $\Lambda_{\rm
QCD}$ (usually taken to be of order $Q$). This assumption is
appropriate to the discussion of physical hadrons, in which confinement
provides a nonperturbative IR cutoff of order $\Lambda_{\rm
QCD}$. However, in perturbative matching calculations of short-distance
coefficients, one usually takes the external quark and gluon states to
be on their mass shells, and, in this situation, soft and collinear gluons
of arbitrarily low energy can be emitted. In order to establish the
consistency of such calculations, one must prove, to all orders in
perturbation theory, that these soft and collinear gluons factor from
the hard-scattering process and that the factorized form is identical to
the conventional one that is obtained in the presence of an infrared
cutoff of order $\Lambda_{\rm QCD}$.\footnote{In
Ref.~\cite{Manohar:2005az}, it was asserted that, if a factorized form
exists when one considers only modes with scales of order $\Lambda_{\rm
QCD}$ or greater, then the short-distance coefficients are independent
of all infrared modes, even in perturbative calculations in which modes
with scales below $\Lambda_{\rm QCD}$ are present. This was shown to be
the case in a one-loop example. However, no general, all-orders proof of
that assertion was given.} In the absence of such a proof, one would
have no guarantee in the matching calculation that low-virtuality soft
and collinear contributions either cancel or can be absorbed entirely into the standard
nonperturbative functions (distribution functions in inclusive processes
and distribution amplitudes in exclusive processes).

At the one-loop level, gluons of arbitrarily low energy can be treated
along with higher-energy soft and collinear gluons, and the conventional
proofs of factorization apply. However, as we shall see, the proof of
factorization of low-energy gluons becomes more complicated beyond one
loop. In multiloop integrals, in the on-shell case, one finds
contributions at leading power in the hard-scattering momentum in which
collinear gluons of low energy couple to soft gluons. Our goal is to 
construct a proof of factorization that takes this possibility into 
account. To  our knowledge, the existing discussions of 
factorization, either in the context of SCET or diagrammatic methods, 
have not addressed this possibility.

In on-shell perturbative calculations in SCET, gluon transverse momenta
extend to zero. Hence, the possibility of low-energy gluons with momenta
collinear to one of the external particles arises. At one-loop level,
the soft and collinear contributions can be separated through the use of
an additional cutoff \cite{Manohar:2006nz}. However, as we have already
mentioned, at two-loop level and higher, a low-energy collinear gluon
can attach to a soft gluon. The SCET action is formulated so that soft
gluons can be decoupled from collinear gluons through a field
redefinition, but there is no corresponding provision to decouple
collinear gluons from soft gluons. Therefore, it seems that, in SCET,
low-energy collinear gluons would be treated most straightforwardly as
part of the soft (or ultrasoft) contribution.  This results in a
factorized form in which the soft function contains gluons with both
soft and collinear momenta and, hence, contains both soft and collinear
divergences.

Alternatively, one can consider a factorized form in which gluons with
collinear momenta are factored completely from the soft function, so
that they reside only in jet functions that are associated with the
initial- or final-state hadrons. Such an alternative factorized form, in
which the soft function is free of collinear divergences, has been
discussed in the context of factorization for the Drell-Yan process in
Refs.~\cite{Bodwin:1984hc,Collins:1985ue,Collins:1989gx}, although the
details of the factorization of gluons with collinear momenta from the
soft function were not given. This alternative factorized form has also
been discussed in connection with resummation of logarithms in, for
example,
Refs.~\cite{Collins:1984ik,Sterman:1986aj,Contopanagos:1996nh,%
Kidonakis:1998bk,Kidonakis:1998nf,Sterman:2002qn}.
Furthermore, it 
has been discussed in an axial gauge in the context of on-shell
quark scattering \cite{Sen:1982bt}. Axial gauges are somewhat
problematic, in that they introduce unphysical singularities into gluon
and ghost propagators. Such singularities could potentially spoil
contour-deformation arguments that are used to ascertain the leading
regions of integration in Feynman diagrams
\cite{Collins:1982wa}.\footnote{For a discussion of a class of gauges
that may ameliorate some of these difficulties, see
Ref.~\cite{Sterman:1978bi}.}  For this reason, we believe that it is
important to construct a proof of factorization in a covariant gauge,
such as the Feynman gauge, which we employ in the present paper.

A factorized form in which the soft function contains no gluons with
collinear momenta has several useful features. One is that contributions
in which there are two logarithms per loop (one collinear and one soft)
reside entirely in the jet functions, which have a diagonal color
structure, rather than in the soft function, which has a more
complicated color structure. Here we focus on a feature that is 
crucial for factorization proofs: A factorized form in which the soft 
function contains no collinear modes allows one to 
establish a cancellation of the soft function when it connects to 
a color-singlet hadron.  As we shall
explain below, if the soft function contains gluons with collinear
momenta, then the cancellation of the soft function fails at leading
order in the large momentum scale.

In this paper, we outline the proof of factorization at leading order
in the hard-scattering momentum for the case of on-shell external
partons. For concreteness, we discuss the example of the exclusive
production of two light mesons in $e^+e^-$ annihilation. In an
intermediate step, the factorized form that we obtain contains a soft
function that is free of collinear divergences. This allows us to
demonstrate the cancellation of the soft function at leading order in the
large momentum scale. Our proof makes use of standard all-orders
diagrammatic methods for proving factorization
\cite{Bodwin:1984hc,Collins:1985ue,Collins:1989gx}. We find that the
factorization of gluons of arbitrarily low energy can be dealt with
conveniently by focusing on the factorization of contributions to loop
integrals from singular regions, {\it i.e.}, regions that contain the
soft and collinear singularities. Such singular regions are discussed
in Refs.~\cite{Collins:1985ue,Collins:1989gx}. However, the coupling 
of low-energy collinear gluons to soft gluons is not discussed in those 
papers.

The remainder of this paper is organized as follows. In
Sec.~\ref{sec:model-amp} we describe the model that we use for the
production amplitude. Section~\ref{sec:leading-regions} contains a
heuristic discussion of the regions of loop momenta that give leading
contributions. This discussion is aimed at making contact with
previous work on factorization and also sets the stage for a more
precise discussion of the singular regions of loop momenta. In
Sec.~\ref{sec:topology}, we discuss the diagrammatic topology of the
leading contributions and also the topology of the soft and collinear
singular contributions. We treat the collinear and soft contributions by
making use of collinear and soft approximations that are valid in the
singular regions. These are discussed in Sec.~\ref{sec:decoupling},
along with the decoupling relations for the collinear and soft singular
contributions. In Sec.~\ref{sec:factorization}, we outline the
factorization of the collinear and soft singularities and describe how
one arrives at the standard factorized form for the production
amplitude. We also outline the proof of factorization in the case in
which the relative momentum between the quark and the antiquark in a meson is
taken to be nonzero. Here, we discuss the difficulty that arises in the
cancellation of the soft function if the soft function contains gluons
with collinear momenta. Finally, in Sec.~\ref{sec:summary}, we summarize
our results.

\section{Model for the amplitude\label{sec:model-amp}}

Let us consider  the exclusive production
of two light mesons in $e^+e^-$ annihilation through a single virtual
photon.
We work in the $e^+e^-$ center-of-momentum
frame and in the Feynman gauge, and we write four-vectors in terms of
light-cone components: $k=(k^+,k^-,\bm{k}_\perp)$, with
$k^{\pm}=(1/\sqrt{2})(k^0\pm k^3)$. We take each meson to be moving in the plus
(minus) direction and to consist of an on-shell quark with momentum $p_{1q}$
($p_{2q}$) and an on-shell antiquark with momentum $p_{1\bar q}$
($p_{2\bar q}$):
\begin{subequations}
\label{p1p2}
\begin{eqnarray}
p_{1q}&=&\left[
\frac{z_1Q}{\sqrt{2}},
\frac{\bm{p}_{1\perp}^2}{\sqrt{2}z_1Q},
\bm{p}_{1\perp}
\right],
\\
p_{1\bar q}&=&\left[
\frac{(1-z_1)Q}{\sqrt{2}},
\frac{\bm{p}_{1\perp}^2}{\sqrt{2}(1-z_1)Q},
-\bm{p}_{1\perp}
\right],
\\
p_{2q}&=&\left[
\frac{\bm{p}_{2\perp}^2}{\sqrt{2}z_2Q},
\frac{z_2Q}{\sqrt{2}},
\bm{p}_{2\perp}
\right],
\\
p_{2\bar q}&=&\left[
\frac{\bm{p}_{2\perp}^2}{\sqrt{2}(1-z_2)Q},
\frac{(1-z_2)Q}{\sqrt{2}},
-\bm{p}_{2\perp}
\right],
\end{eqnarray}
\end{subequations}
where $0<z_i<1$ and $z_i$ does not lie near the endpoints of its range.
The momentum $P_i$ of the meson $M_i$ is given by
\begin{equation}
P_i=p_{iq}+p_{i\bar{q}}.
\end{equation}
The large scale $Q$ is equal to the invariant mass of the virtual
photon, up to corrections of relative order $\bm{p}_{i\perp}/Q$. We
assume that the components of $\bm{p}_{i\perp}$ are all of order
$\Lambda_{\rm QCD}$. It is useful for subsequent discussions to
introduce a dimensionless parameter
\begin{equation}
\label{def-lambda}
\lambda\equiv\Lambda_{\rm QCD}/Q.
\end{equation}
In order to simplify the initial discussion, we
set $\bm{p}_{i\perp}=0$. We will discuss at the end of the
factorization argument the effect of keeping the $\bm{p}_{i\perp}$
nonzero.

\section{Leading regions of loop momenta\label{sec:leading-regions}}

Let us now discuss the regions of loop momenta that are leading in
powers of the large scale $Q$. Our analysis will be somewhat heuristic,
in that, as we will see, the boundaries between the various momentum
types are indistinct. We carry out this analysis in order to make
contact with previous discussions of factorization and to set the stage for
our proof of factorization. That proof focuses on the soft and collinear
singular regions of loop momenta, which {\it are} distinct.\footnote{Power
counting in the neighborhoods of pinch singularities has been discussed
in Refs.~\cite{Sterman:1978bi,Libby:1978bx}.}

Suppose that a virtual gluon with momentum $k$ attaches to external
$q$ or $\bar{q}$ lines with momentum $p_i$ and $p_j$. (In the
remainder of this paper, we call lines that originate in an external $q$
or $\bar{q}$  ``outgoing fermion lines.'') In the limit in which the
components of $k$ are all small compared to the largest components of
$p_i$ and $p_j$, the amplitude associated with this process is 
proportional to
\begin{equation} 
\int d^4k \frac{4p_i\cdot p_j}
{(2p_i\cdot k+i\varepsilon)(-2p_j\cdot k+i\varepsilon)}\frac{1}
{k^2+i\varepsilon}. 
\label{homogeneous}
\end{equation} 
Because the integral is independent of the scale of $k$, leading
contributions arise from arbitrarily small momentum $k$. One can emit an additional
virtual gluon of momentum $k'$ from an outgoing fermion line at a point
to the interior of the emission of a gluon with momentum $k$, provided
that $k'\cdot p_i \gtrsim k\cdot p_i$. Such emissions are arranged in a
hierarchy along the outgoing fermion lines, according to the
virtualities that the emissions produce on the outgoing fermion lines.

Now let us establish some nomenclature to describe the regions of loop
momenta that yield contributions that are leading in powers of the large
scale $Q$. We call such momentum regions ``leading regions.'' We outline
below the construction of an argument to prove that these are the only
possible leading regions. We consider hard ($H$), soft ($S$), 
collinear-to-plus ($C^+$), and collinear-to-minus ($C^-$) momenta, whose 
components have the following orders of magnitude:
\begin{subequations}
\begin{eqnarray}
H\,\,\,&\hbox{: }&Q(1,1,\bm{1}_\perp),
\\
S\,\,\,\,&\hbox{: }&Q\epsilon_S (1,1,\bm{1}_\perp),
\\
C^+&\hbox{: }&Q\epsilon^+[1,(\eta^+)^2,\bm{\eta}^+_\perp],
\\
C^-&\hbox{: }&Q\epsilon^-[(\eta^-)^2,1,\bm{\eta}^-_\perp].
\end{eqnarray}
\end{subequations}
We call a line in a Feynman diagram that carries momentum of type $X$ 
an ``$X$ line.''
The parameters $\epsilon_S$, $\epsilon^+$, and $\epsilon^-$
set the energy scales of the momenta.  We define the soft region of 
momentum space by the condition 
\begin{equation}
\epsilon_S\ll 1.
\label{fact-scale-s}%
\end{equation}
We define the collinear region of momentum space by the conditions 
\begin{eqnarray}
\epsilon^\pm &\lesssim & 1,\nonumber\\
\eta^\pm &\ll &1.
\label{fact-scale-c}%
\end{eqnarray}
In our definitions of momentum regions, the positions of the
boundaries between regions are somewhat vague. That is because 
there is no
clear distinction between the $H$, $S$, and $C^\pm$ regions near the
boundaries between regions: When $\epsilon_S\sim 1$, an $S$ momentum is
essentially an $H$ momentum; when $\eta^\pm\sim 1$, a $C^\pm$ momentum
is essentially an $S$ momentum.

Soft singularities occur in the limit $\epsilon_S\to 0$, and $C^\pm$
singularities occur in the limits $\eta^\pm \to 0$. Hence, we see
that, unlike the soft and collinear momentum regions, the soft and
collinear singularities {\it are} distinct. There are also
singularities that are associated with the scales of the collinear
momenta. These appear in the limit $\epsilon^\pm\to 0$. If $\eta^\pm$
is finite, these are essentially soft singularities, but they can occur
in conjunction with a collinear singularity if $\eta^\pm \to 0$. 
 
We do not consider gluon loop momenta of the ``Glauber'' type
\cite{Bodwin:1981fv}, in which $k^+,k^-\ll|\bm{k}_\perp|$. The reason
for this is that, for exclusive processes, the $k^+$ and $k^-$ contours
of integration are not pinched in the Glauber region, and, hence, one
can always deform them out of that region \cite{Collins:1982wa}. 

If we take $\epsilon^\pm$ to be of order one and $\epsilon_S$ and
$\eta^\pm$ to be of order $\lambda$ [Eq.~(\ref{def-lambda})], then the
resulting momenta are those that are treated  in SCET$_{\rm II}$
\cite{Bauer:2002aj}. Soft momenta with $\epsilon_S$ of order $\lambda^2$
have been considered in Ref.~\cite{Beneke:2000ry} in the context of
two-loop-order contributions to $B$-meson decays, and the possibility of
leading momentum regions involving momenta of arbitrarily small energy
is mentioned in Ref.~\cite{Smirnov:1999bza} for the case of massive
particles.

\begin{widetext}
\begin{center}
\begin{table}
\begin{tabular}{c}
\begin{tabular}{c|ccc}
$k$\,$\backslash$\,$p$ &
   \hspace{7ex}$S$\hspace{7ex}
&\hspace{7ex} $C^\pm$\hspace{7ex}
&\hspace{7ex} $\tilde{C}^\pm$\hspace{7ex}
\\
\hline
\hspace{2ex}$S$\mbox{\hspace{2ex}}&
$\epsilon_{S_k}\sim\epsilon_{S_p}$&
$\epsilon_{p}^{\pm}(\eta^{\pm}_p)^2\lesssim \epsilon_{S_k}\ll
\epsilon_{p}^{\pm}$
&
$\epsilon_{p}^{\pm}\tilde\eta^{\pm}_p\lesssim\epsilon_{S_k}\ll 
\epsilon_{p}^{\pm}
$
\end{tabular}
\\[5ex]
\begin{tabular}{c|cccc}
$k$\,$\backslash$\,$p$ &
 \hspace{7ex}$S$\hspace{7ex}
&\hspace{7ex}$C^\mp$ \hspace{7ex}
&\hspace{7ex}$\tilde{C}^\mp$\hspace{7ex}
&\hspace{7ex}$CC$\mbox{\hspace{7ex}}\\
\hline
\hspace{2ex}$C^\pm$\mbox{\hspace{2ex}}&
$\epsilon^\pm_{k}\sim\epsilon_{S_p}$&
$\epsilon_{p}^\mp(\eta^\mp_{p})^2\lesssim\epsilon^{\pm}_k\lesssim\epsilon^{\mp}_p$
&$\epsilon^\mp_{p}\tilde\eta^{\mp}_{p}\lesssim\epsilon^{\pm}_k\lesssim\epsilon^{\mp}_p$&
$\epsilon^\pm_{k}\sim\epsilon_{{CC}_p}$
\\
$CC$&
$\epsilon_{CC_k}\sim\epsilon_{S_p}$&
$\epsilon_{p}^\mp(\eta^\mp_{p})^2\lesssim\epsilon_{CC_k}\ll\epsilon^{\mp}_p$&
$\epsilon^\mp_{p}\tilde\eta^{\mp}_{p}\lesssim\epsilon_{CC_k}\ll\epsilon^{\mp}_p$&
$\epsilon_{CC_k}\sim\epsilon_{CC_p}$
\end{tabular}
\end{tabular}
\caption{Conditions under which a gluon with momentum $k$ can attach to
a line with momentum $p$. In each table, the left-hand column gives the
momentum type of the gluon with momentum $k$, and the top row gives the
momentum type of the line with momentum $p$. Each entry gives the
conditions that must be fulfilled if the attachment is to satisfy our
conventions for attachments, as described in the text, and also yield a
contribution that is not suppressed by powers of ratios of momentum
components. For purposes of power counting, an $H$ line behaves as a
soft line with $\epsilon_S\sim 1$. The rules for the attachment if
$k$ is a $\tilde{C}^\pm$ momentum are the same as the rules of
attachment if $k$ is a $C^\pm$ momentum. As is explained in the
text, if $k$ is $S$, and the lines to which it attaches have momentum
$p_i$ and $p_j$, then $p_i$ and $p_j$ cannot both be $C^+$ or $C^-$.
Furthermore, if $k$ is $C^\pm$, then at least one of $p_i$ and $p_j$ is
$C^\pm$.
\label{tab:power-counting}}
\end{table}
\end{center}

\end{widetext}

We wish to determine the configurations of the various momentum types in
a Feynman diagram that are leading, in the sense that they are not
suppressed by powers of the ratios of momentum components. 
In our analysis, we begin with the hard subdiagram plus
the bare external $q$ and $\bar q$ for each meson. Then we add one gluon
at a time to the diagram. (Each added gluon possibly contains quark,
gluon, and ghost vacuum polarization loops.) There are many redundant
procedures for adding gluons to obtain a diagram with a given momentum
configuration. We adopt the following convention: We say that a gluon
with momentum $l$ can attach to a line with momentum $p$ only if the
momentum $p+l$ is predominantly of the same type as momentum $p$. For
example, an $S$ gluon with momentum $l$  can attach to a $C^\pm$ line
with momentum $p$ only if $\epsilon_S$ is of order $\epsilon^\pm
\eta^\pm$ or smaller, so that the plus (minus) component of $p+l$ is the
dominant component. Similarly, a $C^\pm$ gluon with momentum $l$ can
attach to an $S$ gluon with momentum $p$ only if $\epsilon^\pm$ is of
order $\epsilon_S$ or smaller, so that all components of $p+l$ are
approximately equal. We call the sum of a $C^\pm$ momentum and an $S$
momentum with $\epsilon_S \sim \epsilon^\pm \eta^\pm$ a $\tilde{C}^\pm$
momentum. The sum of a $C^{\pm}$ momentum and a $C^{\mp}$ momenta with
$\epsilon^\pm(\eta^\pm)^2\ll\epsilon^{\mp}\ll\epsilon^{\pm}$ is also
a $\tilde{C}^{\pm}$ momentum. We also allow the attachment of a $C^\pm$
momentum to a $C^\mp$ momentum with $\epsilon^+\sim \epsilon^-$, and, in
this case, we call the sum of the $C^+$ momentum and $C^-$ momentum a
$CC$ momentum. These combination momenta have the following orders of
magnitude:
\begin{subequations}
\begin{eqnarray}
\tilde{C}^+&\hbox{: }& Q\epsilon^+(1,\tilde{\eta}^+,\bm{\eta}^+_{\perp}),
\\
\tilde{C}^-&\hbox{: }& Q\epsilon^-(\tilde{\eta}^-,1,\bm{\eta}^-_{\perp}),
\\
CC&\hbox{: }& Q\epsilon_{CC}(1,1,\bm{\eta}_{CC\perp}),
\end{eqnarray}
\end{subequations}
where
\begin{equation}
1\gg\tilde{\eta}^\pm \gg (\eta^\pm)^2.
\end{equation}
In order to determine the momenta of attached gluons that can result in
a leading power count, it is useful to consider the
expression~(\ref{homogeneous}).  In the first two factors in the
denominator of the expression~(\ref{homogeneous}), terms of the form
$p_i^2$ and $k^2$ have been dropped. Thus, the denominator of the
expression~(\ref{homogeneous}) gives a lower bound on the order of
magnitude of the exact denominator. Because of our convention for the
allowed momentum types for $k$, the numerator $p_i\cdot p_j$ in the
expression~(\ref{homogeneous}) gives the leading behavior unless
$p_i$ and $p_j$ are both either $C^+$ or $C^-$. For such cases, we need
to consider numerator factors $k^2$, $k\cdot p_i$, and $k\cdot p_j$, in
addition to $p_i\cdot p_j$. Otherwise, we can use the
expression~(\ref{homogeneous}) as it stands to obtain an upper bound
on the magnitude of the factors that appear when one adds a gluon. The
expression~(\ref{homogeneous}) has the useful property that it is
independent of the scales of the momenta $k$, $p_i$, and $p_j$, 
and so it can be used to
determine rules for the leading momentum configurations that are
independent of the scales of the momenta. From these considerations, it
is easy to see that $k$ must be $S$, $C^+$, or $C^-$ in order to obtain
a leading power count. We regard these momentum types as primary, in 
the sense that the loop-integration variables correspond to these 
momenta. Other
momentum types can arise when we add these primary types, following our
convention above for allowed attachments. 
It follows that, if $k$ is
$S$, then $p_i$ and $p_j$ cannot both be $C^+$ or $C^-$. It also
follows that, if $k$ is $C^\pm$, then at least one of $p_i$ and
$p_j$ is $C^\pm$.

If we restore the terms of the form $p_i^2$ and $k^2$ in the
denominators of the expression~(\ref{homogeneous}), then there can be an
additional suppression of the amplitude.\footnote{ In counting powers in
this case, we assume that a
$C^\pm$ line is off shell by an amount of order
$Q^2(\epsilon^\pm)^2(\eta^\pm)^2$ and that an $S$ line is off shell by
an amount of order $Q^2(\epsilon_S)^2$. In the integrations over the
momenta that are associated with the virtual particles, there {\it are}
contributions from the neighborhoods of the mass-shell poles. However,
because the poles in the $k^+$ and $k^-$ complex planes are well
separated, one can always deform the $k^+$ and $k^-$ contours of
integration into the complex plane such that a gluon never has
virtuality smaller than of order the square of its transverse momentum.}
In order to obtain a leading contribution, we must have
\begin{eqnarray}
k\cdot p&\gtrsim& k^2,\nonumber\\
k\cdot p&\gtrsim& p^2.
\label{ksq-psq-conditions}
\end{eqnarray}
Taking into account the additional conditions in
Eq.~(\ref{ksq-psq-conditions}), we obtain the rules for the leading
contributions that are given in Table~\ref{tab:power-counting}. In
Table~\ref{tab:power-counting}, the symbol ``$\sim$'' means that
quantities have the same order of magnitude. In each expression in Table I, 
if the quantity with subscript $k$ is much greater than the quantity with 
subscript $p$, then the
attachment is not allowed because $p+k$ is not essentially of the same
momentum type as $p$. If the quantity with subscript $k$ is much less than 
the quantity with subscript $p$, then the contribution is suppressed
by a power of the ratio of those quantities.\footnote{Suppose that we
add an $S$ gluon to a $C^\pm$ gluon with $\epsilon_S\sim \eta^\pm
\epsilon^\pm$ or that we add a $C^\mp$ gluon to a $C^\pm$ gluon with
$\epsilon^\mp \sim \eta^\pm \epsilon^\pm$. Then, the sum of the momenta
is no longer of the $C^\pm$ type. Because this change in momentum can
propagate through the diagram, such additions of gluons can affect
vertices other than those of the added gluon and propagators other than
those adjacent to a vertex of the added gluon. In these cases, one must
check that the rules in Table~\ref{tab:power-counting} still allow the
attachments at the affected vertices.} The rules in
Table~\ref{tab:power-counting} also apply when the added gluon
attaches to one of the outgoing fermion lines. In that case, one sets
$\eta^+ =0$ or $\eta^- =0$ on the outgoing fermion line. In
Table~\ref{tab:power-counting}, we have not given the rules for the
attachments of gluons with $C^\pm$ or $\tilde C^\pm$ momenta to lines
with $C^\pm$ or $\tilde C^\pm$ momenta. The rules for such attachments
are complicated and cannot be characterized simply in terms of the
magnitudes of the momentum components, as is the case for the
attachments listed in Table~~\ref{tab:power-counting}. For our purposes,
it suffices to note that necessary conditions for such attachments are
given in Eq.~(\ref{ksq-psq-conditions}).

The constraints in Eq.~(\ref{ksq-psq-conditions}) imply that an
attachment of a gluon to a given line is allowed only if the virtuality
that it produces on that line is of order or greater than the virtuality
that is produced by the gluons that attach to that line to the outside
of the attachment in question. Here, and throughout this paper, 
``outside'' means toward the on-shell ends of the external quark and 
antiquark lines. If a gluon with momentum $k$ of type
$C^\pm$, $\tilde{C}^\pm$, $S$, $C^\mp$, or $CC$ attaches to a
$C^\pm$ line from an on-shell outgoing quark or antiquark, it adds virtuality 
$Q^2\epsilon_{k}^\pm (\eta_{k}^\pm)^2$,
$Q^2\epsilon_{k}^\pm \tilde\eta_{k}^\pm$, 
$Q^2\epsilon_{S_k}$, 
$Q^2\epsilon_{k}^\mp$, or 
$Q^2\epsilon_{CC_k}$,
respectively.

\section{Topology of the leading contributions\label{sec:topology}}
\subsection{Topology of the leading momentum regions}

\begin{figure}
\includegraphics[angle=0,height=6.5cm]{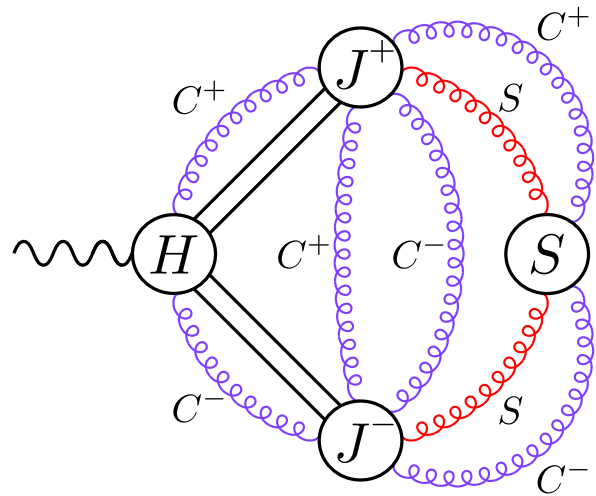}
\caption{ Leading regions for double light-meson production in $e^+e^-$
annihilation. The wavy line represents the virtual
photon.\label{fig:regionsn}
}
\end{figure}

By taking into account the allowed gluon attachments in
Table~\ref{tab:power-counting}, one arrives at the topology of Feynman
diagrams that is shown in Fig.~\ref{fig:regionsn}. This topology is
similar in appearance to topologies that have been discussed previously
in connection with the identification of IR (pinch) singularities in
Feynman diagrams 
\cite{Sterman:1978bi,Sterman:1978bj,Libby:1978bx,Collins:1989gx}.
However, as we will explain, the subdiagrams in Fig.~\ref{fig:regionsn}
contain finite ranges of momenta, whereas those in
Refs.~\cite{Sterman:1978bi,Sterman:1978bj,Collins:1989gx} contain only
infinitesimal neighborhoods of the soft and collinear singularities.
(We will discuss the topology of the soft and collinear singularities
in Sec.~\ref{sec:top-sing}.)

In the topology of Fig.~\ref{fig:regionsn}, there is a jet subdiagram
for each of the collinear regions (corresponding to each light meson),
there is a hard subdiagram that includes the production process at
lowest order in $\alpha_s$, and there is a soft subdiagram. 

We include in the hard subdiagram all propagators that are off shell by
order $Q^2$. That is, we include lines carrying both momentum $H$ and
momentum $CC$ with $\epsilon_{CC}\sim 1$. (The propagators in the Born
process carry momenta $CC$ with $\epsilon_{CC} \sim 1$.) 

The soft subdiagram includes gluons with $S$ momenta, which may contain
quark, gluon, and ghost loops. The soft subdiagram attaches to the jet
subdiagrams through any number of $S$-gluon lines, according to the
rules  in Table~\ref{tab:power-counting}. Note that a gluon carrying momentum $S_i$
cannot attach to a line carrying momentum $S_j$ unless
$\epsilon_{S_i}\sim \epsilon_{S_j}$, and so various part of the soft
subdiagram cannot attach to each other.

The $C^\pm$-jet subdiagram $J^\pm$ contains the external quark lines for
the meson with $C^\pm$ momentum, as well as gluons with $C^\pm$ momenta,
which may contain quark, gluon and ghost loops. We also include in $J^{\pm}$
lines carrying $CC$ momentum with $\epsilon_{CC} \ll 1$ that occur when
a gluon carrying momentum $C^\mp$ from a $J^{\mp}$ jet attaches to a
line carrying $C^\pm$ momentum in $J^{\pm}$. Each jet subdiagram
attaches to the hard subdiagram through the external quark and antiquark
lines and through any number of $C^\pm$ gluons with $\epsilon^\pm \sim
1$. A gluon carrying $C^\pm$ or $\tilde{C}^\pm$ momentum can connect the
$J^\pm$ subdiagram to the $J^\mp$ subdiagram, but only with the
attachments in Table~\ref{tab:power-counting}. Of particular note is the
fact that a gluon carrying momentum $C^\pm$ or $\tilde{C}^\pm$ can
connect a $C^\pm$ jet to an $S$ line in the soft subdiagram, provided
that $\epsilon^\pm \sim \epsilon_S$. This is a feature of scattering
processes in the on-shell case that does not appear when one has an
infrared cutoff of order $\Lambda_{\rm QCD}$. The factorization of
gluons carrying collinear momenta from the soft subdiagram is one of the
principal technical issues that we address in this paper.

In order to prove factorization, we need to show that the
nonperturbative contributions to Feynman diagrams (those with
virtualities of order $\Lambda_{\rm QCD}^2$ or less) either cancel or
can be factored into the meson distribution amplitudes. Specifically, we
will argue that the nonperturbative contributions associated with the
soft divergences factor from the $J^\pm$ subdiagrams and cancel and that
the nonperturbative contributions associated with the $C^\pm$
divergences factor from the $J^\mp$, hard, and soft subdiagrams and can
be absorbed into the $J^\pm$ meson distribution amplitude. These
factorizations and cancellations  establish that the production
amplitude depends only on the properties of the individual mesons, and
not on correlations between the two mesons, except through the hard
subprocess.

\subsection{Two-loop example\label{sec:two-loop}}

In Fig.~\ref{fig:2loopex} we show a two-loop example in which a $C^+$ gluon 
attaches to an $S$ gluon. 
\begin{widetext}
\begin{center}
\begin{figure}
\includegraphics[angle=0,height=7cm]{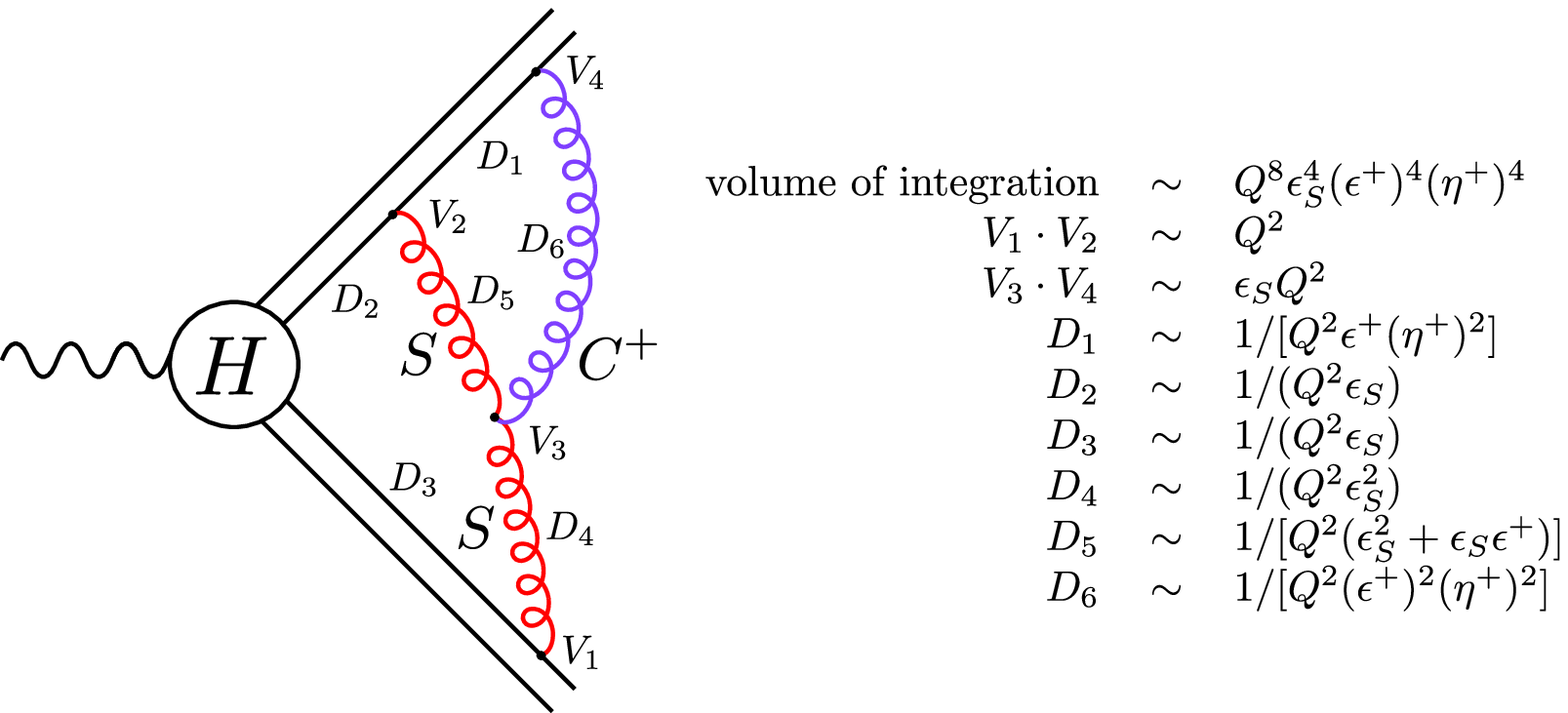}
\caption{A two-loop example in which a $C^+$ gluon attaches to an $S$ 
gluon. The $V_i$ are the vertex factors, and the 
$D_i$ are the propagator factors. \label{fig:2loopex}}
\end{figure}
\end{center}
\end{widetext}
We take the $C^+$ momentum to be
$l_1=Q\epsilon^+(1,\eta^2,\bm{\eta}^+_\perp)$ and the $S$ momentum to be
$l_2=Q\epsilon_S(1,1,\bm{1}_\perp)$. We assume that $\epsilon^+\lesssim
\epsilon_S$, and we route the $l_1$ momentum through the $D_5$
propagator. Then, we find the factors for the diagram that are shown
on the right side of Fig.~\ref{fig:2loopex}. Combining these
factors, we obtain the following order of magnitude for the two-loop
correction: $\epsilon_S\epsilon^+/(\epsilon_S^2+\epsilon_S\epsilon^+)$.
We see that this result is independent of $Q$, as expected, and is also
independent of $\eta$. This contribution is leading if
$\epsilon^+\sim\epsilon_S$, but it vanishes in the limit
$\epsilon^+/\epsilon_{S}\to 0$, in accordance with the rule in
Table~\ref{tab:power-counting}.

\subsection{Topology of the singular momentum regions\label{sec:top-sing}}

In the preceding discussion, as we have noted, the soft and collinear
momentum regions are not well distinguished. If $\eta^\pm \sim 1$,
then a collinear momentum is virtually identical to a soft momentum.
Similarly, if the components of a soft momentum have significantly
different sizes, then a soft momentum can be virtually identical to a 
collinear momentum. In
discussions of factorization, we rely on collinear approximations that
are accurate only for $\eta^\pm \ll 1$. In order to
apply such approximations, we must avoid the problems in distinguishing
soft and collinear momenta that arise near the boundaries between these
regions. Furthermore, the soft approximation for the attachment of a
soft gluon to a $C^\pm$ line becomes inaccurate as the soft momentum 
becomes more nearly a $C^\pm$ momentum. Again, we encounter a problem 
that occurs near the
boundary between momentum regions. In the discussion that follows, we
avoid such boundary issues by focusing on infinitesimal neighborhoods
of the soft and collinear singularities (singular regions).\footnote{It
has been suggested that problems that arise near boundaries between
momentum regions can be avoided by implementing a subtraction scheme
that is akin to the Bogoliubov-Parasiuk-Hepp-Zimmerman formalism for 
subtraction of ultraviolet
divergences \cite{Collins:1985ue,Collins:1989gx}. Such a subtraction
scheme has not yet been constructed, although one-loop examples have
been given in the context of the zero-bin-subtraction method of SCET
\cite{Manohar:2006nz}.} As a first step in proving factorization, we
will demonstrate the factorization of these singular regions.

The topologies of soft and collinear singular regions have been
discussed in the context of factorization theorems for inclusive
processes in Refs.~\cite{Collins:1985ue,Collins:1989gx}. These
topologies follow from the rules for power counting that we have given 
in Sec.~\ref{sec:leading-regions}.
Let us describe the relationship of the topologies of the singular
regions to the topologies in Fig.~\ref{fig:regionsn}. The $C^\pm$
singularities reside in the outermost part of the $J^\pm$ subdiagram,
which we call the $\tilde{J}^\pm$ subdiagram. (We consider the
$\tilde{J}^\pm$ subdiagram to be part of the $J^\pm$ subdiagram, and we
call the part of the $J^\pm$ subdiagram that excludes the
$\tilde{J}^\pm$ subdiagram the $J^\pm-\tilde{J}^\pm$ subdiagram.) The
soft singularities reside in the outermost part of the $S$ subdiagram,
which we call the $\tilde{S}$ subdiagram. (We consider the $\tilde{S}$
subdiagram to be part of the $S$ subdiagram, and we call the part of the
$S$ subdiagram that excludes the $\tilde{S}$ subdiagram the
$S-\tilde{S}$ subdiagram.) $S$ singular gluons connect the $\tilde{S}$
subdiagram only to the $\tilde{J}^\pm$ subdiagrams. The $\tilde{J}^\pm$
subdiagrams connect to the $J^\pm$, $J^\mp$, $S$, and $H$ subdiagrams
via $C^\pm$ gluons. We emphasize that the $\tilde{J}^\pm$ subdiagrams
connect, via $C^\pm$ gluons to the $\tilde{S}$ subdiagram. This last
type of connection is a feature that was not included in the discussion
of leading (pinch) singularities in
Refs.~\cite{Collins:1985ue,Collins:1989gx}. Otherwise, the topologies
that we find are the same as in
Refs.~\cite{Collins:1985ue,Collins:1989gx}, provided that we identify
the hard subdiagram in those references with the union of all of the
subdiagrams in our topology except for $\tilde{S}$, $\tilde{J}^+$, and
$\tilde{J}^-$. We call this union $\tilde{H}$.

\section{Collinear and soft approximations and decoupling relations%
\label{sec:decoupling}}

Our strategy is to show that contributions from the $\tilde{J}^\pm$
subdiagrams factor from the $J^\mp$, hard, and $S$ subdiagrams and can
be absorbed into the $J^\pm$ meson distribution amplitude and that
contributions from the $\tilde{S}$ subdiagram factor from the
$\tilde{J}^\pm$ subdiagrams and cancel. We treat the contributions
from the $C^\pm$ singular regions by making use of a collinear-to-plus
(minus) approximation \cite{Bodwin:1984hc,Collins:1985ue,Collins:1989gx}
for the $C^\pm$ gluons that attach the $\tilde{J}^\pm$ subdiagram to the
$J^\mp$, $H$, and $\tilde S$ subdiagrams. The $C^\pm$ approximations
capture all of the collinear-to-plus (minus) singularities, but
become increasingly inaccurate as one moves away from the
singularities. Similarly, we treat the contributions from the $S$
singular regions by using a soft approximation for the gluons with $S$
momentum that attach the $\tilde{S}$ subdiagram to the $\tilde{J}^\pm$
subdiagrams. The soft approximation captures all of the soft
singularities, but becomes increasingly inaccurate as one moves away
from the singularities.

\subsection{Collinear approximation}

Let us now describe the collinear approximation explicitly. Suppose that
a gluon carrying momentum in the $C^\pm$ singular regions attaches to a
line carrying $H$, $C^\mp$, $\tilde{C}^\mp$, $S$, or $CC$ momentum.
Then, we can apply a collinear approximation to that gluon
\cite{Bodwin:1984hc,Collins:1985ue,Collins:1989gx} with no loss of
accuracy. The collinear-to-plus ($C^+$) and collinear-to-minus
($C^-$) approximations consist of the following replacements in the 
gluon-propagator numerator:
\begin{equation}
g_{\mu\nu}
\,\,\Longrightarrow\,\,
\left\{
\begin{array}{ll}
\displaystyle
\frac{k_\mu \bar n_{1\nu}}{k\cdot \bar n_1-i\varepsilon}& (C^+),
\\[2ex]
\displaystyle
\frac{k_\mu \bar n_{2\nu}}{k\cdot \bar n_2+i\varepsilon}& (C^-).
\end{array}
\right.
\end{equation}
The index $\mu$ corresponds to the attachment of the gluon to the hard,
soft, or $J^\mp$ subdiagram, and the index $\nu$ corresponds to the
attachment of the gluon to the $J^\pm$ subdiagram. Our convention is
that $k$ flows out of a $C^+$ line and into a $C^-$ line. There is
considerable freedom in choosing the auxiliary vectors $\bar n_1$ and
$\bar n_2$. In order to reproduce the amplitude in the collinear
singular region, it is only necessary to have $\bar n_1\cdot p_{1q}> 0$
(or $\bar n_1\cdot p_{1\bar{q}}> 0$)
and 
$\bar n_2\cdot p_{2q} > 0$ 
(or $\bar n_2\cdot p_{2\bar{q}} > 0$). 
We choose $\bar n_1$ and $\bar n_2$ to
be lightlike vectors in the minus and plus directions such that, for
any vector $q$, $q\cdot \bar n_1=q^+$ and $q\cdot \bar n_2=q^-$. 
The $C^\pm$ approximation relies on the fact that the $\pm$ component of
$k$ dominates in the collinear limit, provided that the $\mu$ index
connects to a current in which the $\mp$ component is nonzero. 
Because of this last stipulation, we cannot apply the collinear
approximations to a gluon carrying momentum in the $C^\pm$ singular
region when it attaches to a line that is also carrying momentum in the
$C^\pm$ singular region. In the $C^\pm$ approximation, the gluon's
polarization is longitudinal, {\it i.e.}, proportional to the gluon's
momentum, which is essential to the application of graphical Ward
identities to derive decoupling relations. 

\subsection{Soft approximation}

Suppose that a gluon that carries momentum $k$ in the $S$ singular
region attaches to a line carrying momentum $p$ that lies outside the $S$ 
singular region. Then we can apply the soft approximation without
loss of accuracy. The soft approximation
\cite{Grammer:1973db,Collins:1981uk} consists of replacing
$g_{\mu\nu}$ in the gluon-propagator numerator with $k_\mu p_\nu/k\cdot
p$, where the index $\mu$ corresponds to the attachment of the gluon to
the line with momentum $p$. Unlike the collinear approximation, the soft
approximation depends on the momentum of the line to which the gluon
attaches. For the attachment of the gluon with momentum $k$ to any line
with momentum in the $C^+$ ($C^-$) singular region, the soft
approximation consists of the following replacements in the
gluon-propagator numerator:
\begin{equation}
g_{\mu\nu}
\,\,\Longrightarrow\,\,
\left\{
\begin{array}{ll}
\displaystyle
\frac{k_\mu n_{1\nu}}{k\cdot n_1+i\varepsilon}& (C^+),
\\[2ex]
\displaystyle
\frac{k_\mu n_{2\nu}}{k\cdot n_2-i\varepsilon}& (C^-),
\end{array}
\right.
\end{equation}
where $n_1$ and $n_2$ are lightlike vectors that are proportional
to $p_{1q}$ (or $p_{1\bar{q}}$) and $p_{2q}$ (or $p_{2\bar{q}}$),
respectively, and are normalized such that, for any vector $q$,
$n_1\cdot q=q^-$ and $n_2\cdot q=q^+$. The index $\mu$ contracts into
the line carrying the momentum of type $C^+$ ($C^-$).

\subsection{Decoupling relations}

\begin{figure} 
\includegraphics[angle=0,width=8.5cm]{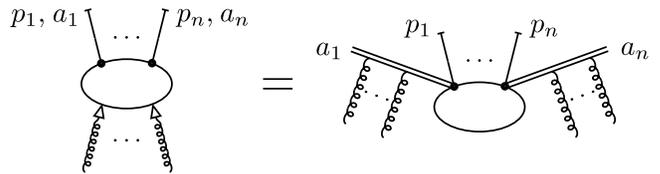}
\caption{Graphical form of the decoupling relations for collinear and
soft gluons. The relations show the decoupling of longitudinally
polarized gluons, which are represented by curly lines. The $C^+$ ($C^-$)
decoupling relation applies when the longitudinally polarized gluons all
have momenta in the $C^+$ ($C^-$) singular region. The $S^+$ ($S^-$)
decoupling relation applies when the longitudinally polarized gluons all
have momenta in the $S$ singular region and the subdiagram that is
represented by an oval contains only lines with momenta in the
$C^+$ ($C^-$) singular region. The longitudinally polarized gluons are
to be attached in all possible ways to the oval. The arrows on the gluon
lines represent the factors $k^\mu \bar n^\nu/(k\cdot \bar n)$ [$k^\mu
n^\nu/(k\cdot n)$] that appear in the collinear (soft) approximation.
The external lines with hash marks are truncated. In addition, the
subdiagram can include any number of untruncated on-shell external legs
(not shown), provided that the polarizations of the on-shell gluons
are orthogonal to their momenta. $p_i$ are momenta, and the $a_i$ are
color indices. The double lines are $C^+$, $C^-$, $S^+$, or $S^-$
eikonal lines, which are described in the text. \label{fig:wi}}
\end{figure}

Once we have implemented a collinear or soft approximation, we can make
use of decoupling relations to factor contributions to the
amplitude. The decoupling relations for collinear and soft gluons have
the same graphical form, which is shown in Fig.~\ref{fig:wi}.

If any number of longitudinally polarized gluons carrying momenta in
the $C^+$ ($C^-$) singular region attach in all possible ways to a
subdiagram, then the $C^+$ ($C^-$) decoupling relation applies. The
subdiagram can have any number of truncated external legs and any number
of untruncated on-shell external legs, provided that the polarization of 
each on-shell gluon is orthogonal to its momentum.
In the $C^+$ ($C^-$) case, the eikonal
(double) lines shown in Fig.~\ref{fig:wi} have the Feynman rules that
a vertex is $\mp igT_a \bar n_{1\mu}$ ($\pm igT_a \bar n_{2\mu}$) and a
propagator is $i/(k\cdot \bar n_1-i\varepsilon)$ [$i/(k\cdot \bar
n_2+i\varepsilon)$], where the upper (lower) sign in the vertex is for
eikonal lines that attach to quark (antiquark) lines.  Here, $T_a$ is an
$SU(3)$ color matrix in the fundamental representation. (Our convention
is that a QCD gluon-quark vertex is $igT_a\gamma_\mu$.) We call these
eikonal lines $C^+$ and $C^-$ eikonal lines, respectively.

An analogous decoupling relation holds when any number of longitudinally
polarized gluons with momenta in the soft singular region attach in all
possible ways to a subdiagram that contains only lines with momenta in
the $C^+$ ($C^-$) singular regions. Again, the subdiagram can have any
number of truncated external legs and any number of untruncated on-shell
external legs, provided that the polarization of each untruncated
on-shell gluon is orthogonal to its momentum. In this case, the
eikonal lines have the Feynman rules that a vertex is $\pm igT_a 
n_{1\mu}$ ($\mp igT_a n_{2\mu}$) and a propagator is $i/(k\cdot
n_1+i\varepsilon)$ [($i/(k\cdot  n_2-i\varepsilon)$] when the subdiagram
is $C^+$ ($C^-$). We call these eikonal lines $S^+$ and $S^-$ eikonal
lines, respectively.\footnote{The decoupling relations rely on the fact
that, in the collinear and soft singular regions, the momenta of the
attached gluons are effectively parallel to each other. This fact is
obvious in the case of the collinear singular regions. In the case of
the soft singular region, this is also the case because the currents to
which the soft gluons attach are all in the plus (minus) direction when
the soft gluons attach to a $C^+$ ($C^-$) subdiagram. Hence, only the
minus (plus) components of the gluons' momenta appear in invariants. In
Refs.~\cite{Collins:1985ue,Collins:1989gx}, an alternative definition of
the soft approximation is given in which this fact is made manifest. In
this definition, if the soft gluon attaches to the $\tilde{J}^+$
($\tilde{J}^-$) subdiagram, then the momentum $k$ is replaced, in the
subdiagram and in the soft approximation, with a collinear momentum
$\tilde{k}=\bar{n}_1 k\cdot n_1$ ($\tilde{k}=\bar{n}_1 k\cdot n_1$).
This alternative definition of the soft approximation is equivalent to
the one that is implied by the Feynman rules for SCET. It has the
property that the decoupling relation (field redefinition in SCET) holds
even outside the soft singular region.}

The Feynman rules for the eikonal lines in the collinear and soft
decoupling relations are summarized in Table~\ref{tab:cs-eikonal}.
\begin{table}
\begin{center}
\begin{tabular}{c|cc}
Type&
Vertex&Propagator\\
\hline
&&\\[-1.5ex]
$C^+$&
\hspace{2ex}
$\mp i g T_a \bar{n}_{1\mu}$
\hspace{2ex}
&
\hspace{2ex}
$\displaystyle \frac{i}
                                   {k\cdot \bar{n}_1-i\varepsilon}$ 
\hspace{2ex}
\\[2.5ex]
$C^-$&
\hspace{2ex}
$\pm i g T_a \bar{n}_{2\mu}$
\hspace{2ex}
&
\hspace{2ex}
$\displaystyle \frac{i}
                                   {k\cdot \bar{n}_2+i\varepsilon}$ 
\hspace{2ex}
\\[2.5ex]
$S^+$&
\hspace{2ex}
$\pm i g T_a n_{1\mu}$
\hspace{2ex}
&
\hspace{2ex}
$\displaystyle \frac{i}
                                   {k\cdot {n}_1+i\varepsilon}$ 
\hspace{2ex}
\\[2.5ex]
$S^-$&
\hspace{2ex}
$\mp i g T_a n_{2\mu}$
\hspace{2ex}
&
\hspace{2ex}
$\displaystyle \frac{i}
                                   {k\cdot {n}_2-i\varepsilon}$ 
\hspace{2ex}
\end{tabular}
\caption{
Feynman rules for the collinear 
($C^\pm$) and soft ($S^\pm$)  eikonal lines. 
The upper (lower) sign is for 
the eikonal line that attaches to a quark (antiquark) line.
\label{tab:cs-eikonal}}
\end{center}
\end{table}
\section{Factorization\label{sec:factorization}}

Now let us describe the factorization of the contributions from the
$C^+$, $C^-$, and $S$ singular regions. 

We can determine the momentum assignments that give singular
contributions by making use of the power-counting rules that we have
outlined in Sec.~\ref{sec:leading-regions}. When we apply these rules to
the attachments of gluons with momenta in the singular regions, the
symbol $\sim$ and the phrase ``of the same order'' mean that quantities
do not differ by an infinite factor, while the phrases ``much less than''
and ``much greater than'' mean that quantities {\it do} differ by an
infinite factor. Hence, for gluons with momenta in the singular regions,
our convention that an allowed attachment of a gluon cannot change the
essential nature of the momentum of the line to which it attaches has
the following meaning: The attaching gluon cannot have an energy that is
greater by an infinite factor than the energy of the line to which it
attaches.

The rules in Sec.~\ref{sec:leading-regions} lead to complicated
relationships between the allowed momenta of gluons in a given
diagrammatic topology. However, there is a general principle, which we
have already mentioned, that allows us to organize the discussion: The 
attachments of gluons to a given line must be ordered so that a given
attachment produces a virtuality along the line that is of order or greater 
than the virtualities that are produced by the attachments that lie to 
the outside of it. In particular, the virtuality that a $C^\pm$,
$\tilde{C}^\pm$, or $S$ singular gluon produces on a $C^\mp$,
$\tilde{C}^\mp$, or $S$ line is of order its energy times the energy of
the line to which it attaches.

\subsection{Characterization of the singular contributions}

The relationships between allowed momenta lead to a hierarchy of scales
as the singular limits are approached. Consider, for example, the 
contribution in which an additional soft gluon is attached to 
the diagram of Fig.~\ref{fig:2loopex} to the same outgoing fermion lines as 
the other gluons, but to the outside of them. In order for this 
contribution to be leading, the additional soft gluon must produce a 
virtuality on the outgoing fermion lines that is of order or less than 
the virtuality of $D_1$ or $D_{3}$. The former condition implies that the 
energy scale of the additional soft gluon $\epsilon_S'$ must be of order 
or less than $\epsilon^+(\eta^+)^2$. Since $\epsilon_S\sim \epsilon^+$, 
this implies that $\epsilon_S'\sim \epsilon_S (\eta^+)^2$. That is, in 
the collinear singular limit, $\epsilon_S'$ is infinitesimal with 
respect to $\epsilon_S$.

From such arguments it is clear that an infinite hierarchy of
virtualities of various infinitesimal orders appears. However, these
orders of virtuality are well separated in the singular limits.
That is, the various gluon energy scales differ by infinite factors,
as in our example. This property allows us to organize the singular
contributions in such a way that we can apply the soft and $C^\pm$
approximations to obtain the factorized form.

In order to carry out the factorization, we need to distinguish two
cases for the ordering of the energy scale of a collinear momentum
relative to the energy scale of a soft momentum. Both of these orderings
can yield contributions that are nonvanishing in the limits
$\epsilon_S\to 0$, $\eta^\pm\to 0$.

{\bf Case 1:} As $\epsilon_S\to 0$, $\epsilon^\pm/\epsilon_S$ is finite.
(It is easy to see that the contribution in which
$\epsilon^\pm/\epsilon_S$ goes to zero vanishes. See for example,
Sec.~\ref{sec:two-loop}.) In this case, we say that the collinear
singular momentum and the soft singular momentum have energies that are
of the same order.

{\bf Case 2:} As $\epsilon_S\to 0$, $\epsilon_S/\epsilon^\pm\to
0$.\footnote{This is the situation that was discussed in
Refs.~\cite{Collins:1985ue,Collins:1989gx}.} In this case we say that
the soft singular momentum has energy that is infinitesimal in
comparison with the energy of the collinear singular momentum.

We will use an iterative procedure to factor gluons at the different
levels of the hierarchy of energy scales. It is useful to establish
first a general nomenclature to characterize this hierarchy of energy
scales. We characterize each level in the hierarchy by the energy
scale of the soft singular gluons in that level. We call that energy
scale the ``nominal scale''. We call soft singular and collinear
singular gluons that have energies of order this scale nominal-scale
gluons. We call collinear singular gluons that have energies that are
infinitely larger than the nominal energy scale but infinitely smaller
than the next-larger soft-gluon scale ``large-scale'' collinear gluons.
The nominal-scale collinear gluons are of the type in case~1 above with
respect to the nominal-scale soft gluons. The large-scale  collinear
gluons are of the type in case~2 above with respect to the nominal-scale
soft gluons.

\subsection{Factorization of the singular 
contributions\label{sec:fact-sing}}

Let us now describe the factorization of the singular contributions.
We make use of an iterative procedure in which gluons of higher
energies are factored before gluons of lower energies. As we shall see,
this ordering of the factorization procedure is convenient because it
allows us to apply the decoupling relations rather straightforwardly to
decouple gluons whose attachments lie toward the inside of the Feynman
diagrams before we decouple gluons whose attachments lie to the outside
of the Feynman diagrams.

\begin{figure}
\includegraphics[angle=0,height=7cm]{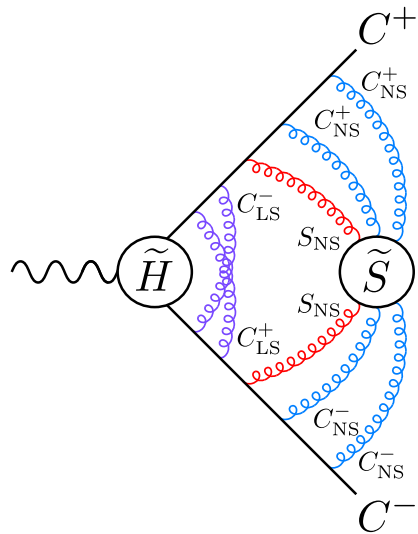}
\caption{Diagram to illustrate the factorization of large-scale
collinear gluons and nominal-scale soft and collinear gluons for double
light-meson production in $e^+e^-$ annihilation. $C^i_{\textrm{LS}}$
denotes a large-scale $C^i$ singular gluon, $C^i_{\textrm{NS}}$ denotes
a nominal-scale $C^i$ singular gluon, and $S_{\textrm{NS}}$ denotes a
nominal-scale $S$ singular gluon. 
\label{fig:fact1}%
}
\end{figure}

We will illustrate the factorization of the large-scale collinear gluons
and the nominal-scale soft and collinear gluons for double light-meson
production in $e^+e^-$ annihilation by referring to the diagram that is
shown in Fig.~\ref{fig:fact1}. In this diagram, we have suppressed
gluons with energies that are much less than the nominal scale. These
gluons have attachments that lie to the outside of the attachments of
the gluons that are shown explicitly. In the diagram in
Fig.~\ref{fig:fact1}, each gluon represents any finite number of gluons,
including zero gluons. For clarity, we have suppressed the antiquark
lines in each meson and we have shown explicitly only the attachments of
the gluons to the quark line in each meson and only a particular
ordering of those attachments. However, we take the diagram in
Fig.~\ref{fig:fact1} to represent a sum of many diagrams, which includes
all of the attachments that we specify in the arguments below of the
singular gluons to the quark and antiquark in each meson, to other
singular gluons, and to the $\tilde{H}$ subdiagram.

\subsubsection{Factorization of the large-scale $C^\pm$ 
gluons\label{sec:large-c}}

We begin with the large-scale $C^\pm$ gluons that have the largest
energy scale, and proceed iteratively through all of the scales of the
large-scale $C^\pm$ gluons. In the first step of the iteration,
those are gluons with finite-energy collinear singular momenta. In the
subsequent steps, only gluons with infinitesimal collinear
singular momenta are present. Gluons with relatively infinitesimal
energies may attach to a gluon that carries a $C^\pm$ singular momentum.
We still consider that gluon to carry $C^\pm$ singular momentum.

First, we wish to apply the $C^+$ approximation and the $C^+$
decoupling relation (Fig.~\ref{fig:wi}) to decouple the large-scale
$C^+$ gluons that originate in the $\tilde{J}^+$ subdiagram from the
$\tilde{H}$ and $\tilde{J}^-$ subdiagrams. In applying the decoupling
relation, we need to know the extent of the subdiagram in
Fig.~\ref{fig:wi}: Eikonal lines appear at the points at which the
subdiagram is truncated. 

We include the attachments of gluons with large-scale $C^+$ momenta to the
$\tilde{J}^-$ subdiagram that are allowed by our conventions and by
power counting. Here, and in the discussions to follow, we consider a
$C^\pm$ gluon to be attached to the $\tilde{J}^\mp$ subdiagram if and
only if its momentum routes through $\tilde{H}$.

We include {\it all} of the attachments of gluons with large-scale $C^+$
momenta to $\tilde{H}$. We include the attachments that are allowed
by our conventions and by power counting. However, we also include
formally attachments to $\tilde{H}$ that yield vanishing contributions
in the singular limits. (In subsequent iterations, we include formally,
as well, the vanishing attachments of large-scale $C^+$ gluons to points
on $C^-$ eikonal lines that lie to the interior of the outermost
attachment of a $C^-$ singular gluon.) 

In applying the $C^+$ decoupling relation, we do not include attachments
of  gluons with large-scale $C^+$ momentum to a gluon with
nominal-scale $S$ momentum: Such attachments violate our convention
for allowed attachments because they alter the nature of the $S$
singular momentum. However, as we mentioned above, gluons with
nominal-scale $S$ singular momenta can attach to a gluon with
large-scale $C^+$ singular momentum without altering the nature of
the $C^+$ singular momentum. We carry these attachments along as we
attach the gluon with finite $C^+$ singular momentum to other lines in
the diagram. We follow this same procedure in discussions below in
treating gluons whose energies are infinitesimal with respect to the
energy of an $S$ or a $C^\pm$ singular gluon to which they attach.

The allowed attachments of gluons with large-scale $C^+$ momenta to a
$C^-$ singular line lie to the inside of the attachments of gluons with
nominal-scale $S$ or $C^\pm$ momenta. Therefore, one might expect that,
when the $C^+$ decoupling relation is applied, a $C^+$ eikonal-line
contribution would appear at the vertex immediately to the outside of
the outermost allowed attachment of a large-scale $C^+$ gluon. In fact,
such an eikonal-line contribution vanishes because the propagator on the
$C^-$ singular line just to the outside of the outermost allowed
attachment of a gluon with large-scale $C^+$ singular momentum is on
shell, and, in the case of a gluon line, has physical polarization
(polarization orthogonal to its momentum), up to relative corrections of
infinitesimal size. Therefore, we omit such eikonal-line contributions
in applying the decoupling relation.

Then, the result of applying the decoupling relation is that the gluons
with large-scale $C^+$ momenta attach to $C^+$ eikonal lines
that attach to the outgoing fermion lines in $\tilde{J}^+$ just to the
outside of the $\tilde{H}$ subdiagram.

Next we decouple the gluons that originate in the $\tilde{J}^-$
subdiagram and have large-scale $C^-$ momenta from the $\tilde{H}$
and $J^+$ subdiagrams. The procedure follows the same argument as
for the gluons with large-scale $C^+$ singular momenta, except for
one new ingredient: We must include formally the vanishing attachments
of the gluons with large-scale $C^-$ momenta to the $C^+$ eikonal
lines from the previous step. Note that we need to include only the
attachments that lie to the interior of the attachment of the outermost
gluon with $C^+$ singular momentum in order to apply the $C^-$
decoupling relation. The result of applying the $C^-$ decoupling
relation is that gluons with large-scale $C^-$ momenta attach to
$C^-$ eikonal lines that attach to $C^-$ outgoing fermion lines just to the
outside of the $\tilde{H}$ subdiagram.

\begin{figure}
\includegraphics[angle=0,height=7cm]{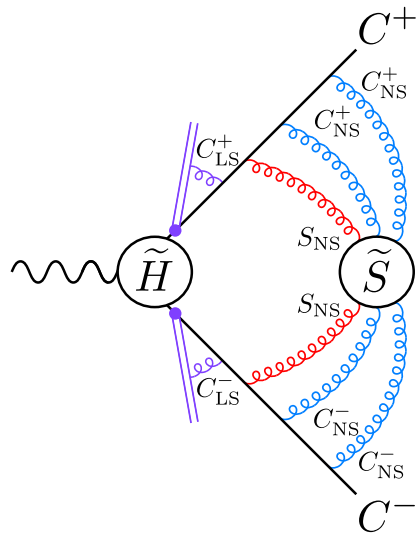}
\caption{Diagram representing the sum of diagrams
that occurs after one applies the decoupling of the large-scale
collinear gluons that is described in
Sec.~\ref{sec:large-c}.
\label{fig:fact2}%
}
\end{figure}

Now, we iterate this procedure for the large-scale $C^\pm$ gluons at
the next-lower energy scale. The result of applying the $C^\pm$
decoupling relations is that the large-scale $C^\pm$ gluons attach to
$C^\pm$ eikonal lines that attach to the outgoing fermion lines just to
the inside of the $C^\pm$ eikonal lines from the previous iteration. It
is easy to see that, on each outgoing fermion line, the $C^\pm$ eikonal
line from the current iteration can be combined with the $C^\pm$ eikonal
line from the previous iteration into a single $C^\pm$ eikonal line. On
the combined $C^\pm$ eikonal line, the attachments of $C^\pm$ gluons
with the smaller energy scale lie to the outside of the attachments of
gluons with the larger energy scale. (Other orderings yield vanishing
contributions.) We continue iteratively in this fashion until we have
factored all of the large-scale $C^\pm$ gluons. After this decoupling
step, the sum of diagrams represented by Fig.~\ref{fig:fact1} becomes a
sum of diagrams represented by Fig.~\ref{fig:fact2}.

\subsubsection{Factorization of the nominal-scale $C^\pm$ 
gluons\label{sec:nom-c}}

\begin{figure}
\includegraphics[angle=0,height=7cm]{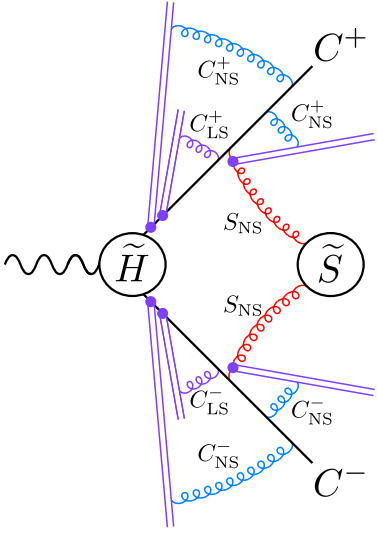}
\caption{Diagram representing the sum of diagrams
that occurs after one applies the initial decoupling of the
nominal-scale collinear gluons that is described in
Sec.~\ref{sec:nom-c}.
\label{fig:fact3}%
}
\end{figure}

Next we factor the nominal-scale $C^\pm$ gluons. In applying the $C^+$
decoupling relation, we include the allowed attachments of these gluons
to the $\tilde{J}^-$ subdiagram and the attachments to the nominal-scale
soft gluons. We also include formally the vanishing contributions from
the attachments of the nominal-scale gluons to the $\tilde{H}$
subdiagram and to the $C^-$ eikonal lines. Because of the ordering
of virtualities along a line with $C^+$ singular momentum, the outermost
attachment to such a line of a gluon with nominal-scale $C^+$ momentum
must lie to the outside of the outermost attachment of a gluon with
nominal-scale $S$ momentum. It is then easy to see that, for every
attachment described above of a $C^+$ line to a line with
momentum that is not $C^+$ singular, the $C^+$ approximation holds
exactly. The $C^-$ propagator that lies to the outside of the outermost 
allowed attachment of a gluon with
nominal-scale $C^+$ momentum to a line with $C^-$ singular momentum is
on-shell, and, in the case of a gluon line, has physical polarization,
up to relative corrections of infinitesimal size. Therefore, when we
apply the $C^+$ decoupling relation, no eikonal line appears at the
vertex immediately to the outside of this outermost attachment of a
gluon with nominal-scale $C^+$ momenta. The result of applying the $C^+$
decoupling relation is that the nominal-scale $C^+$ gluons attach to
several $C^+$ eikonal lines. These eikonal lines attach in the following
locations: to the outgoing $C^+$ fermion lines just to the outside of
$\tilde{H}$, but to the inside of the large-scale $C^+$ eikonal lines;
just to the soft-gluon side of each vertex involving a nominal-scale
soft gluon and a $C^+$ singular gluon of the large scale or a larger
scale. In a similar fashion, we factor the nominal-scale $C^-$
gluons. The result of applying the $C^-$ decoupling relation is that the
$C^-$ singular gluons attach to several $C^-$ eikonal lines. These
eikonal lines attach to the following locations: to the outgoing $C^-$
fermion lines just to the outside of $\tilde{H}$, but to the inside of
the large-scale $C^-$ eikonal lines; just to the soft-gluon side of each
vertex involving a nominal-scale soft gluon and a $C^-$ singular gluon
of the large scale or a larger scale. After this decoupling
step, the sum of diagrams represented by Fig.~\ref{fig:fact2} becomes
the sum of diagrams represented by Fig.~\ref{fig:fact3}.

\subsubsection{Factorization of the nominal-scale $S$ 
gluons\label{sec:nom-s}}

\begin{figure}
\includegraphics[angle=0,height=7cm]{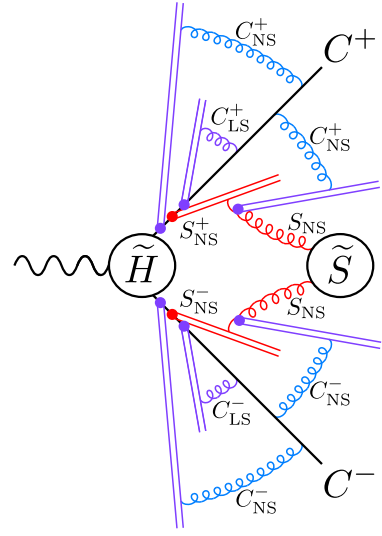}
\caption{Diagram representing the sum of diagrams
that occurs after one applies the decoupling of the nominal-scale soft
gluons that is described in Sec.~\ref{sec:nom-s}.
\label{fig:fact4}%
}
\end{figure}

We now wish to apply the soft decoupling relations to factor the
nominal-scale soft gluons. In order to do this, we implement the $S^\pm$
approximations for the attachments of the soft gluons to the $C^\pm$
singular lines of the large scale or a larger scale. However, we
make a slight modification to the soft approximation by combining the
momentum of the nominal-scale soft gluon with the total momentum of the
associated nominal-scale $C^\pm$ eikonal line. Then, when we implement
the $S^\pm$ decoupling relations, the nominal-scale $C^\pm$ eikonal
lines are carried along with the nominal-scale soft-gluon attachments.
In applying the $S^+$ decoupling relation, we include attachments of
nominal-scale soft gluons to the $\tilde{J}^+$ subdiagram, and in
applying the $S^-$ decoupling relation, we include attachments of
nominal-scale soft gluons to the $\tilde{J}^-$ subdiagram. Because we
have already factored the attachments of nominal-scale $C^\pm$
gluons, the $S^\pm$ approximations hold, up to relative corrections of
infinitesimal size. We also include vanishing attachments of the
nominal-scale soft gluons to the large-scale eikonal lines
\cite{Collins:1985ue}, including only those soft-gluon attachments
that lie to the inside of the outermost $C^+$-gluon attachments. The
$C^\pm$ propagator that lies to the outside of the outermost allowed
attachment of a nominal-scale soft gluon to a $C^\pm$ line is
on shell, up to relative corrections of infinitesimal size. Therefore,
when we apply the $S^\pm$ decoupling relations, no $S^\pm$ eikonal lines
appear at the vertices just to the outside of the outermost allowed
attachments. The result of applying the $S^\pm$ decoupling relations
is that soft gluons attach to $S^\pm$ eikonal lines. These eikonal lines
attach to the outgoing $C^\pm$ fermion lines just to the outside of the
nominal-scale $C^{\pm}$ eikonal lines  and just to the inside of the
large-scale $C^\pm$ eikonal lines. Associated with each attachment
of a nominal-scale soft gluon to an $S^\pm$ eikonal line is a $C^\pm$
eikonal line to which nominal-scale $C^\pm$ gluons attach. After this
decoupling step, the sum of diagrams represented by Fig.~\ref{fig:fact3}
becomes a sum of diagrams represented by Fig.~\ref{fig:fact4}.

\subsubsection{Further factorization of the nominal-scale $C^\pm$ 
gluons\label{sec:nom-c-f}}

\begin{figure}
\includegraphics[angle=0,height=7cm]{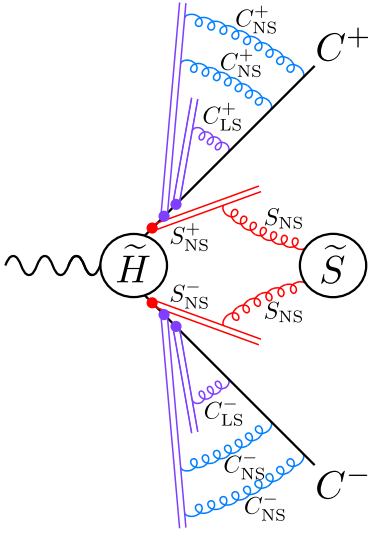}
\caption{Diagram representing the sum of diagrams
that occurs after one applies the further decoupling of the
nominal-scale collinear gluons that is described in
Sec.~\ref{sec:nom-c-f}.
\label{fig:fact5}%
}
\end{figure}

\begin{figure}
\includegraphics[angle=0,height=7cm]{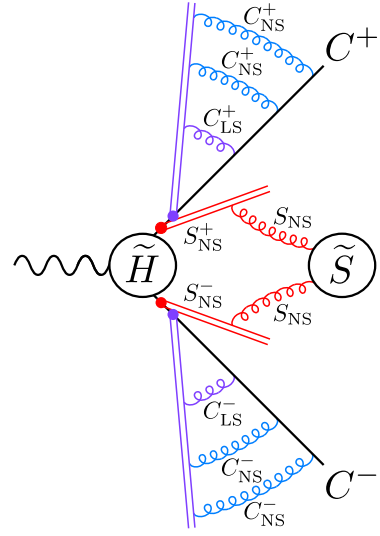}
\caption{Diagram representing the sum of diagrams
that occurs after one completely decouples the large-scale collinear
gluons and the nominal-scale soft and collinear gluons.
\label{fig:fact6}%
}
\end{figure}

We next factor the nominal-scale $C^\pm$ gluons from the $S^\pm$ eikonal
lines. In order do this, we include formally the vanishing contributions
that arise when one attaches the nominal-scale $C^\pm$ gluons to all
points on the $S^\pm$ eikonal lines that lie to the inside of the
outermost attachment of a nominal-scale soft gluon. We also make use
of the following facts: Each nominal-scale $C^\pm$ eikonal line that
attaches to an outgoing $C^\pm$ fermion line is identical to the eikonal
line that one would obtain by applying the $C^\pm$ decoupling relation
to the attachments of the nominal-scale $C^\pm$ gluons to an on-shell
fermion line; each nominal-scale $C^\pm$ eikonal line that attaches
to a nominal-scale gluon is identical to the eikonal line that one would
obtain by applying the $C^\pm$ decoupling relation to the attachments of
nominal-scale $C^\pm$ gluon to an on-shell gluon line. Then, applying
the $C^+$ decoupling relation, we find that the nominal-scale $C^+$
gluons attach to $C^+$ eikonal lines that attach to the outgoing fermion
lines just to the inside of the large-scale $C^+$ eikonal lines.
Similarly, applying the $C^-$ decoupling relation, we find that the
nominal-scale $C^-$ gluons attach to $C^-$ eikonal lines that attach to
the outgoing fermion lines just to the inside of the large-scale
$C^-$ eikonal lines.  This result is represented by 
the diagram that is shown in Fig.~\ref{fig:fact5}. 
The nominal-scale $C^\pm$ eikonal lines can then
be combined with the large-scale $C^\pm$ eikonal lines. After
performing those steps, we arrive at the final factorized form,
which is represented by the diagram in Fig.~\ref{fig:fact6}.
\subsubsection{Completion of the factorization}

Now we can iterate the procedure that we have given in
Secs.~\ref{sec:large-c}--\ref{sec:nom-c-f}, taking the nominal scale to
be the next-smaller soft-gluon scale. In these subsequent iterations,
we include formally, in the steps of Secs.~\ref{sec:large-c} and
\ref{sec:nom-c}, the vanishing contributions from the attachments of the
large-scale and nominal-scale $C^+$ and $C^-$ gluons to the soft gluons
of higher levels and to the $S^+$ and $S^-$ eikonal lines that are
associated with those soft gluons. (We also include formally the
vanishing contributions from the attachments of the large-scale and
nominal-scale $C^+$ and $C^-$ gluons to $\tilde{H}$, as in the first
iteration.)

Proceeding iteratively through all of the soft-gluon scales, we produce
new nominal-scale $S^\pm$ eikonal lines at each step that
attach to the outgoing fermion lines just to the outside of the
existing $S^\pm$ eikonal lines. The nominal-scale $C^\pm$ eikonal
lines that attach to the outgoing $C^\pm$ fermion lines after the steps
of Sec.~\ref{sec:nom-c} are situated just to the inside of these
nominal-scale $S^\pm$ eikonal lines. After the further factorization
of the nominal-scale $C^\pm$ gluons that is described in
Sec.~\ref{sec:nom-c-f}, the $S^\pm$ eikonal lines that attach to a given
outgoing fermion can be combined into a single $S^\pm$ eikonal line. The
soft gluons of a lower energy scale attach to the outside of the soft
gluons of a higher energy scale. This is the only ordering that
produces a nonvanishing contribution.

\begin{figure}
\includegraphics[angle=0,height=7cm]{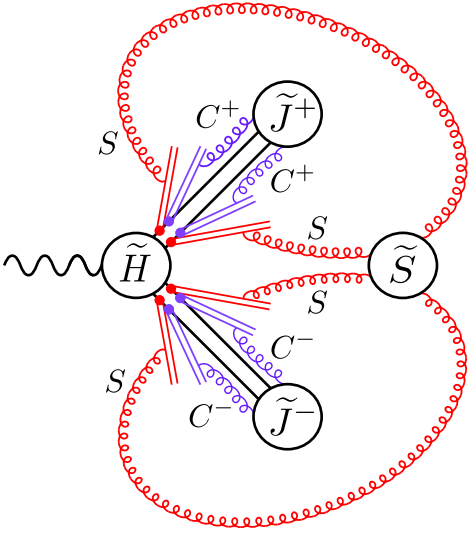}
\caption{Illustration of the factorized form for double light-meson
production in $e^+e^-$ annihilation. After the use of the decoupling
relations, gluons with momenta in the $S$ singular region attach to
$S^{\pm}$ eikonal lines and gluons with momenta in the $C^{\pm}$
singular regions attach to $C^{\pm}$ eikonal lines.\label{fig:softfa}%
}
\end{figure}

Following this procedure, we arrive at the standard factorized form for
the singular contributions. The $\tilde{S}$ subdiagram now attaches only
to $S^+$ eikonal lines that attach to the outgoing fermion lines from
$\tilde{J}^+$ just outside of $\tilde{H}$ and to $S^-$ eikonal lines
that attach to the outgoing fermion lines from $\tilde{J}^-$ just
outside of $\tilde{H}$. The attachments involve only gluons with $S$
singular momenta. All of the $C^\pm$ singular contributions are
contained in the $J^\pm$ subdiagram, which attaches via $C^\pm$ singular
gluons to $C^\pm$ eikonal lines that attach to the outgoing fermion
lines from $\tilde{J}^\pm$ just outside of the $S^\pm$ eikonal lines.
This factorized form is illustrated in Fig.~\ref{fig:softfa}.

\subsection{Cancellation of the eikonal lines\label{sec:eikonal-cancel}}

At this point the $\tilde{S}$ subdiagram and associated
soft eikonal lines, which we call $\bar S$, have the form of the
vacuum-expectation value of a time-ordered product of four eikonal lines:
\begin{eqnarray}
\bar{S}(x_{1 q},x_{1\bar q},x_{2 q},x_{2\bar q})&=&\nonumber\\
&&\hspace{-8em}\langle 0\vert T\{[x_{1\bar q},\infty^+][\infty^+,x_{1 q}]
\otimes
[x_{2\bar q},\infty^-][\infty^-,x_{2 q}]\} \vert 0 \rangle_S,\nonumber\\
\label{S-tilde-me}
\end{eqnarray}
where 
\begin{equation}
[x,y]=\exp\left[\int_{x}^y igT_a A_\mu^a dx^\mu\right] 
\end{equation}
is the exponentiated line integral (eikonal line) running between $x$ and
$y$, $\infty^+=(\infty,0,\bm{0}_\perp)$, and
$\infty^-=(0,\infty,\bm{0}_\perp)$. The symbol $\otimes$ indicates a
direct product of the color factors that are associated with the
soft-gluon attachments to meson~1 and the soft-gluon attachments to
meson~2. We note that eikonal-line self-energy subdiagrams, which were
absent in our derivation of $\bar S$, vanish for lightlike eikonal
lines in the Feynman gauge. The subscript on the matrix element
indicates that only contributions from the soft singular region are
kept.

Because the $H$ and $J^+-\tilde{J}^+$ subdiagrams are
insensitive to a momentum in the $S$ singular region flowing through
them, we can ignore the difference between $x_{1 q}$ and $x_{1\bar q}$
in Eq.~(\ref{S-tilde-me}). Then the $S^+$ eikonal lines cancel. Note
that this cancellation relies on the color-singlet nature of the
external meson. In a similar fashion, we can ignore the difference
between $x_{2q}$ and $x_{2\bar q}$ in Eq.~(\ref{S-tilde-me}), and the $S^-$
quark and antiquark eikonal lines cancel. 

We can make a Fierz rearrangement to decouple the color factors of
the $\tilde{J}^{+}$ and $\tilde{J}^{-}$ subdiagrams from
$\tilde{H}$. Then,  we can write the $\tilde{J}^{\pm}$ subdiagrams
and their associated eikonal lines, which we call $\bar{J}^{+}$ and
$\bar{J}^{-}$, as follows: 
\begin{widetext}
\begin{eqnarray}\label{light-cone-dist2}%
\bar{J}^\pm_{\alpha\beta}(z_i)
& = & \frac{P_i^\pm}{\pi}\int_{-\infty}^{+\infty} d x^\mp
\exp[
-
i(2z_i-1)P_i^\pm x^\mp] 
\langle  M_i(P_i)\vert \bar{\Psi}_\alpha(x^\mp) 
T\{[x^\mp,\infty^{\mp} ]
 [\infty^\mp ,-x^\mp]\}\Psi_\beta(-x^\mp)\vert 
0\rangle_{C^\pm}.
\end{eqnarray}
Here, $z_i$ is the fraction of $P_i^\pm$ that is carried by the quark in
meson~$i$, $\alpha$ and $\beta$ are Dirac indices, and the upper (lower)
sign in Eq.~(\ref{light-cone-dist2}) corresponds to $i=1$ ($i=2$). It is
understood that the fields $\Psi$ and $\bar{\Psi}$ in the matrix element
are in a color-singlet state. The subscripts on the matrix elements
indicate that only the contributions from the collinear singular
regions are kept.

There is a partial cancellation of the eikonal lines in $\bar{J}^+$
and $\bar{J}^-$, with the result that the residual eikonal lines run
directly from $-x^\mp$ to $x^\mp$: 
\begin{eqnarray}\label{light-cone-dist3}%
\bar{J}^\pm_{\alpha\beta}(z_i)
 =  \frac{P_i^\pm}{\pi}\int_{-\infty}^{+\infty} d x^\mp
\exp[
-
i(2z_{i}-1)P_i^\pm x^\mp] 
\langle M_i(P_i)\vert \bar{\Psi}_\alpha(x^\mp) 
P[x^\mp,-x^\mp]
\Psi_\beta(-x^\mp)\vert 
 0\rangle_{C^\pm}.
\end{eqnarray}
\end{widetext}
Here, we have written the time-ordered product of the exponentiated line
integral as a path-ordered product. Because the integrations over $z_1$
and $z_2$ have nonvanishing ranges of support in $\tilde{H}$, $x^\mp$
and $-x^\mp$ in Eq.~(\ref{light-cone-dist3}) are 
typically separated by a distance of order $1/Q$.
This shows that the $C^\pm$ singular
contributions that have energies much less than $Q$ cancel, once they
have been factored.

\subsection{Factorized form}

We have shown that the contributions from $C^\pm$ singular regions
factor from the $\tilde{J}^\mp$, $S$, and $H$ subdiagrams and are
contained entirely in the $\bar{J}^\pm$ subdiagrams and that the
contributions from the $S$ singular region factor from the
$\tilde{J}^\pm$ subdiagrams and cancel. The $\bar{J}^\pm$
subdiagrams each have precisely the form of a meson distribution
amplitude. Hence, we have arrived at the conventional factorized form,
except for the following facts: the $\bar{J}^\pm$ subdiagrams contain
only the infinitesimal $C^\pm$ singular regions, whereas they are
conventionally defined to contain finite regions of integration; the
$\tilde{H}$ subdiagram is not yet free of nonperturbative contributions
from collinear momenta with transverse components of order $\Lambda_{\rm
QCD}$ or less.\footnote{At this stage, we have shown that, if one uses
dimensional regularization for the soft and collinear divergences in the
production amplitudes, then the soft poles in $\epsilon=(4-d)/2$ cancel
and the collinear poles can be factored into $\bar{J}^\pm$.}

Next we extend the ranges of integration in the logarithmically
ultraviolet divergent integrals in $\bar{J}^\pm$ from infinitesimal
neighborhoods of the collinear singularities to finite neighborhoods
that are defined by an ultraviolet cutoff $\mu_F\sim Q$, which is the
factorization scale. In making such an extension, we do not encounter
any new singularities in $\bar{J}^\pm$. The soft singularities that do
not involve the eikonal lines have already been shown to cancel. There
is the possibility that $S$ or $C^\mp$ singularities could arise from
the eikonal lines in $\bar{J}^\pm$. However, as we have mentioned, after
the cancellation of the quark and antiquark eikonal lines, the remaining
segment of eikonal line is finite in length, with length of order $1/Q$.
Hence, $S$ or $C^\mp$ modes with virtualities much less than $Q$ cannot
propagate on these eikonal lines.

Finally, having extended the momentum ranges in $\bar{J}^\pm$, we
redefine $\tilde{H}$ to be the factor that, when convolved with
$\bar{J}^\pm$, produces the complete production amplitude. This is
precisely the conventional definition of the hard subdiagram. Since the
soft divergences have canceled and the collinear divergences are
contained in $\bar{J}^\pm$, $\tilde{H}$ is a finite function (after
ultraviolet renormalization), and depends only on the scales $Q$ and
$\mu_F$ and the renormalization scale. Therefore, $\tilde{H}$
contains only contributions from momenta of order $Q$ or $\mu_F$,
{\it i.e.}, from momenta in the perturbative regime. We have now
established the conventional factorized form
for the production amplitude $\mathcal{A}$, which reads
\begin{equation}
\mathcal{A}=\bar{J}^-\otimes\tilde{H}\otimes\bar{J}^+,
\end{equation}
where the symbol $\otimes$ denotes a convolution over the longitudinal
momentum fraction $z_i$ of the corresponding meson and 
we have suppressed the Dirac indices on $\bar{J}^\pm$ and $\tilde{H}$. 
The $\bar{J}^{\pm}$ are now given by Eq.~(\ref{light-cone-dist3}), but 
without the subscript $C^\pm$ on the matrix element. They can be 
decomposed into a sum of products of Dirac-matrix and kinematic factors 
and the standard light-cone distributions for the mesons.

\subsection{Nonzero relative momentum between the quark and antiquark}

Now let us return to the situation in which the $\bm{p}_{i\perp}$ in
Eq.~(\ref{p1p2}) are nonzero and of order $\Lambda_{\rm QCD}$. In this
case the outgoing quark and antiquark in each meson are moving in
slightly different light-cone directions. Therefore, we must define
separate singular regions $C_{1q}$ ($C_{2q}$) for the quark direction
and $C_{1\bar q}$ ($C_{2\bar q}$) for the antiquark direction in meson
1~(2). 

As we have mentioned, in defining the collinear approximations, we
can choose any auxiliary vectors $\bar n_i$ that, for the collinear
singular region associated with $p_i$, satisfy the relation $\bar n_i
\cdot p_i>0$. We choose the lightlike auxiliary vector $\bar n_1$,
which is in the minus direction, for both the $C_{1q}$ and $C_{1\bar
q}$ singular regions and the lightlike auxiliary vector $\bar n_2$,
which is in the plus direction, for both the $C_{2q}$ and $C_{2\bar
q}$ singular regions. That is, we take the same collinear approximation
for the collinear regions associated with the quark and the antiquark in
a meson. Then the factorization of the collinear singular regions goes
through exactly as in the case $p_\perp=0$. 

For gluons with momenta in the soft singular region, one can still
define soft approximations, but the approximations are different for the
couplings to lines in the quark and antiquark collinear singular
regions.  When gluons with momenta in the soft singular region
attach to lines with momenta in the $C_{iq}$ ($C_{i\bar q}$)
singular region, one can use a unit lightlike vector $n_{iq}$
($n_{i\bar q}$) that is proportional to $p_{iq}$ ($p_{i\bar q}$) to
define the soft approximation. Then, the gluons with momenta in the soft
singular region still factor. However, in the factored form, the soft
eikonal line that attaches to the quark (antiquark) line in meson $i$ is
parametrized by the auxiliary vector $n_{iq}$ ($n_{i\bar q}$). Because
$n_{iq}$ and $n_{i\bar q}$ differ by an amount of relative order
$\lambda$ [Eq.~(\ref{def-lambda})], the quark and antiquark soft eikonal
lines in each meson fail to cancel completely. These noncancelling soft
contributions violate factorization because they couple one meson to the
other in the production amplitude.

If we take the approximation $n_{iq} = n_{i\bar{q}}$, but keep
$n_{jq}\neq n_{j\bar q}$, then the quark and antiquark eikonal lines
cancel in meson~$i$, but not in meson~$j$. However, the remaining soft
subdiagram, which attaches only to the quark and antiquark eikonal
lines in $\bar{J}$, can be absorbed into the
definition of the $\bar{J}$ subdiagram for meson~$j$.\footnote{It can
be shown, by making use of the methods in Sec.~\ref{sec:fact-sing}, that
the configuration in which the soft subdiagram attaches only to the
quark and antiquark soft eikonal lines that are associated with
$\bar{J}^+$ ($\bar{J}^-$) is precisely the configuration that one would
obtain by using the soft decoupling relation to factor a soft subdiagram
that attaches only to $\bar{J}^+$ ($\bar{J}^-$). Here one must use the
fact that the contributions in which $C^+$ ($C^-$) collinear gluons with
infinitesimal energy attach to the collinear eikonal line in $\bar{J}^+$
($\bar{J}^-$) vanish, owing to the finite length of the eikonal line.}
Therefore, we see that we obtain a violation of factorization only if
the quark and antiquark eikonal lines fail to cancel in both mesons.
Hence, the violations of factorization that arise from the soft
function are of relative order $\lambda^2$. 

In order to express the amplitude in terms of the light-cone
distributions $\bar J^\pm$ in Eq.~(\ref{light-cone-dist2}), it is 
necessary to neglect in $\tilde{H}$ the minus and transverse components of
$p_{1q}$ and $p_{1\bar q}$ and the plus and transverse components of
$p_{2q}$ and $p_{2\bar q}$. In doing so, we  make an error of
relative order $p_{i\perp}/Q\sim \lambda$. 

\subsection{Failure of the soft cancellation for low-energy collinear 
gluons}

Now let us discuss the cancellation of the soft diagram for the
factorized form in which the soft subdiagram contains collinear gluons.
As we have mentioned, such a factorized form is the one that would seem
to follow most straightforwardly from SCET
\cite{Bauer:2001yt,Bauer:2002nz}.
 Suppose that a gluon with momentum $k$ attaches to the soft eikonal
line that attaches to the quark line in meson~1. That contribution
contains a factor $1/k\cdot n_{1q}$, which is singular in the limit in
which $k$ becomes collinear to $p_{1q}$ ($n_{1q}$). On the other hand,
for the contribution in which the gluon with momenta $k$ attaches to the
soft eikonal line that attaches to the antiquark line in meson~1, there
is a factor $1/k\cdot n_{1\bar q}$, which is not singular in the limit in
which $k$ becomes collinear to $p_{1q}$. Hence, the attachments of the
gluon with momentum $k$ to the quark and antiquark lines fail to cancel
when $k$ is in the $C_{1q}$ (or $C_{1\bar q}$) singular region.
Furthermore, the uncanceled contribution is not suppressed by a power of
$Q$ and is, in fact, divergent. Thus, we see that, in the factorized
form in which the soft subdiagram contains collinear gluons, the soft
subdiagram fails to cancel, and one cannot establish the conventional
factorized form.\footnote{There is also a potential
difficulty in apply the soft approximation to low-energy collinear
gluons. For example, as we have mentioned, if a soft gluon attaches to
the $J^+$ subdiagram, then soft approximations in SCET and
Refs.~\cite{Collins:1985ue,Collins:1989gx} and the soft decoupling
relation involve the replacement of the soft momentum $k$ with a
collinear momentum $\tilde{k}=\bar{n}_1 k\cdot n_1$ in the $\tilde{J}^+$
subdiagram. Hence, $\tilde{k}$ vanishes when $k$ becomes $C^+$.}. It
might seem that one could recover the cancellation of the soft
subdiagram by setting $p_{i\perp}$ exactly to zero. However, at
$p_\perp=0$, the cancellation of the quark and antiquark eikonal lines
becomes ill-defined for $k$ collinear to the quark and antiquark because
of the infinite factors that arise from the eikonal denominators $k\cdot
n_{1q}$ and $k\cdot n_{1\bar{q}}$. \footnote{One might also consider
the possibility of defining a single soft-approximation auxiliary vector
$n_i$ for each meson, where $n_i$ lies between $n_{iq}$ and $n_{i\bar
q}$. However, the resulting soft approximation fails to reproduce the
collinear divergences that occur if the soft subdiagram contains
low-energy gluons that are parallel to the quark or the antiquark.} We
note that this issue also arises in inclusive processes, for example the
Drell-Yan process, in the decoupling of the soft subdiagram from
color-singlet hadrons.

\section{Summary\label{sec:summary}}

We have established, to all orders in perturbation theory, factorization
of the amplitude for the exclusive production of two light mesons in
$e^+e^-$ annihilation through a single virtual photon for the case in
which the external mesons are represented by an on-shell quark and an
on-shell antiquark. The case of on-shell external particles is important
for perturbative matching calculations.

The presence of on-shell external particles opens the possibility of
soft and collinear momentum modes of arbitrarily low energy. In this
situation, low-energy collinear gluons can couple to soft gluons. That
coupling leads to additional complications in the factorization proof.
Nevertheless, we have shown that one can derive the standard factorized
form, in which the production amplitude is written as a hard factor
convolved with a distribution amplitude for each meson. The hard factor
is free of soft and collinear divergences and depends only on the
hard-scattering scale $Q$, the collinear factorization scale $\mu_{C}$,
and an ultraviolet renormalization scale. The meson distribution
amplitudes contain all of the collinear divergences and all of the
nonperturbative contributions that involve virtualities of order
$\Lambda_{\rm QCD}$ or less. We find that the factorization formula
holds up to corrections of relative order $\Lambda_{\rm QCD}/Q$.

As an intermediate step in the factorization proof, we obtain a form in 
which the soft
subdiagram does not contain gluons with momenta in the collinear
singular regions. This form of factorization may be useful in the
resummation of soft logarithms, as the contributions with two logarithms
per loop are contained entirely in the jet functions, which are diagonal
in color. It is essential in establishing the standard factorized form
for exclusive processes with on-shell external partons because, as we
have shown, the cancellation of the attachment of the soft diagram to a
color-singlet hadron fails at leading order in $Q$ if the soft
subdiagram would contain gluons with momenta that are collinear to the
constituents of the hadron. This issue also arises in inclusive
processes in the decoupling of the soft subdiagram from color-singlet
hadrons.

In on-shell perturbative calculations in SCET, low-energy gluons with
momenta collinear to the external particles can appear. At two-loop
level and higher, these low-energy collinear gluons can couple to soft
gluons. Since SCET has no provision to decouple the collinear gluons
from the soft gluons, it seems that it would be most straightforward in
SCET to treat the low-energy gluons as part of the soft contribution. In
such an approach, the soft subdiagram contains gluons with momenta in
both the soft and collinear singular regions. As we have said, the soft
subdiagram would fail to cancel in this case, and one would not
achieve the standard factorized form. Therefore, in the absence of a
further factorization argument, there would be no assurance in a
matching calculation that the low-virtuality contributions could all be
absorbed into the meson distribution amplitudes: Some low-virtuality
contributions might be associated with a soft function that could
not be factored from the  meson distribution amplitudes.

Alternatively, one could abandon the notion that SCET should reproduce
the contributions of full QCD on a diagram-by-diagram basis and assume
that SCET is valid only after one sums over all Feynman diagrams.
Furthermore, one could consider the collinear action in SCET to apply
to all collinear momenta of arbitrarily low energy. Then, as is asserted
in Ref.~\cite{Bauer:2002nz}, the production amplitude in SCET would take
the form of a hard-scattering diagram, a $\bar{J}^+$ light-cone
distribution and a $\bar{J}^-$ light-cone distribution that are
convolved with the hard subdiagram, and a soft subdiagram that is free
of collinear momenta and that connects to the $\bar{J}^+$ and
$\bar{J}^-$ light-cone distributions with interactions that are given by
the collinear action. That factorized form is the one that we would
obtain after the decoupling of the collinear gluons from the soft gluons
if we were to extend the ranges of integration in the $\tilde{S}$,
$\bar{J}^+$, and $\bar{J}^-$ subdiagrams from the singular
regions to finite regions of $S$, $C^+$ and $C^-$ momenta. Issues of
double counting arise when one extends the ranges of integration. They
could be dealt with, for example, by making use of the method of zero-bin
subtractions \cite{Manohar:2006nz}. Once the double-counting issues are
resolved, our proof shows that such a form for the amplitude is
correct. However, this result does not follow obviously from QCD
or from SCET. It requires a derivation, such as the one that we have
given in this paper.

The low-energy contributions that we have discussed involve integrands
that are homogeneous in the integration momenta. Therefore, one might
argue that, if one applies the method of regions \cite{Beneke:1997zp}, 
then such contributions lead to scaleless integrals and vanish. The
difficulty in making use of such an argument to prove factorization is
that the method of regions extends the range of integration for each
region to infinity. There is no proof of the validity of such an
extension, and, hence, there is the possibility of double counting.
Double counting between the soft and collinear subdiagrams is dealt with
in SCET through the use of zero-bin subtractions \cite{Manohar:2006nz}.
However, the zero-bin subtractions are formulated rigorously in terms of
a hard cutoff. In Ref.~\cite{Manohar:2006nz}, examples of the zero-bin
subtraction in dimensional regularization in one-loop perturbation
theory are given. To our knowledge, no proof of an
all-orders zero-bin subtraction scheme in dimensional regularization has
been given.

In physical hadrons, gluon momenta are cut off by confinement at a
scale of order $\Lambda_{\rm   QCD}$. In that situation, one does not
need to consider the possibility that collinear gluons can attach to soft
gluons in order to demonstrate the factorization of nonperturbative
contributions, {\it i.e.}, those contributions that involve momentum
components of order $\Lambda_{\rm   QCD}$. However, if one wishes to
factor logarithmic contributions up to a scale of order $Q$, for
example, for the purpose of resummation, then it is again necessary to
treat the attachments of collinear gluons to soft gluons along the lines
that we have described in this paper.

In this paper, we have focused on a specific exclusive process. However,
we expect that our method can be generalized straightforwardly to the
other exclusive processes and, possibly, to inclusive processes. In the
latter case, one must consider Glauber-type momenta explicitly, as
contributions that arise from such momenta cancel only once one has
summed over all possible final-state cuts
\cite{Bodwin:1984hc,Collins:1983ju,Collins:1985ue,Collins:1988ig,Collins:1989gx}.
However, it seems plausible that one can implement this cancellation,
using standard techniques, independently of the factorization arguments
that we have presented here.

\begin{acknowledgments}
We thank John Collins and George Sterman for many useful comments and
suggestions. We also thank Thomas Becher, Dave Soper, and Iain Stewart
for helpful discussions. We thank In-Chol Kim for his assistance in
preparing the figures in this paper. The work of G.T.B. and X.G.T. was
supported in part by the U.S. Department of Energy, Division of High
Energy Physics, under Contract No.~DE-AC02-06CH11357. The research of
X.G.T.\ was also supported by Science and Engineering Research Canada.
The work of J.L.\ was supported by the Korea Ministry of Education,
Science, and Technology through the National Research Foundation under
Contract No.~2010-0000144.
\end{acknowledgments}

\end{document}